\documentclass[twocolumn,showpacs,amsfonts,aps,prc,floatfix,nofootinbib,superscriptaddress,groupedaddress]{revtex4-1}
\usepackage{bbm}
\usepackage{amsmath,epsfig,amssymb,subfigure,bm,dsfont}
\usepackage{lipsum}
\usepackage{float}  
\usepackage{array}
\usepackage{booktabs}
\usepackage{makecell}
\usepackage[colorlinks,linkcolor=red,citecolor=blue]{hyperref}
\usepackage{amsmath}
\usepackage{mathrsfs}
\usepackage{amssymb}
\usepackage{graphicx,epsfig,latexsym,overpic,amssymb,color}
\usepackage{multirow}
\usepackage[section]{placeins}

\preprint{\today}
\begin{document}
\title{Probing the refined performance of the Categorical-Boosting algorithm to the Hartree-Fock-Bogoliubov mass model with different Skyrme forces}
\author{Jin-Liang Guo}
\affiliation{School of Physics, Zhengzhou
University, Zhengzhou 450001, China}
\author{Hua-Lei
Wang}\email{wanghualei@zzu.edu.cn} \affiliation{School of Physics, Zhengzhou University, Zhengzhou 450001, 
China}

\author{Zhen-Zhen Zhang}
\affiliation{School of Physics, Zhengzhou
University, Zhengzhou 450001, China}
\author{Min-Liang Liu}
\affiliation{Key Laboratory of High Precision Nuclear Spectroscopy,
Institute of Modern Physics, Chinese Academy of Sciences, Lanzhou
730000, China} \affiliation{School of Nuclear Science and
Technology, University of Chinese Academy of Sciences, Beijing
100049, China}

\begin{abstract}
\label{abstract}
Nuclear mass can offer profound insights into many physical branches, e.g., nuclear physics and astrophysics, while the predicted accuracy by nuclear mass models is usually far from satisfactory until now, especially within the fully microscopic self-consistent mean-field theory. In this project, we present the predictive power for the binding energy within the the Hartree-Fock-Bogoliubov (HFB) methods with six widely used Skyrme forces (SkM*, SkP, SLy4, SV-min, UNEDF0 and UNEDF1) and evaluate the refined performance of the machine learning based on a novel Categorical Boosting (CatBoost) algorithm to the Skyrme HFB mass models. The root-mean-square (rms) deviations between the bare HFB calculations with different Skyrme forces and the available experimental data range from the minimum, about 1.43 MeV, for the UNEDF0 parameter set to the maximum, about 7.03 MeV, for the SkM* paraterer set. For the CatBoost-refined HFB predictions, the predictive power can be significantly improved. All the prediction accurancies on the testing set can reach the level around 0.2 MeV and, meanwhile, the large model bias can be reduced. The model-repair coefficients for the adopted Skyrme parameter sets are uniformly more than 80\%. Moreover, for 21 newly measured nuclei outside AME2020, the predicted masses by the CatBoost-refined HFB models are also in good agreement with the experimental data, 
illustrating their good generalization abilities. Intrestingly, it is found that the optimal Skyrme parameter set that possesses the highest predictive power for the bare HFB mass calculations may be not the best candidate for the CatBoost-refined HFB model, indicating the different abilities of picking up the missing ``physics'' for different Skyrme forces by the CatBoost algorithm.  

%
\end{abstract}
%
\maketitle
%
\section{Introduction}
\label{Introduction}
The importance of the masses (equivalently, binding energies) of nuclei has been widely stressed, e.g., for nuclear physics and astrophysics, in the literature~\cite{RevModPhys.75.1021,RevModPhys.29.547}. Both experimentally and thoretically, exploring the nuclear masses is still of particular interest though tremendous progress has so far been made~\cite{PhysRevC.101.045204}. For instance, about 3000 nuclei have been found or synthesized experimentally and around 2500 nuclei have been measured accurately~\cite{Wang_2021}. Depending on different theoretical models~\cite{Weizsacker1935,PhysRevLett.108.052501,WANG2014215,10.1143/PTP.113.785}, it is predicted there should exists a more vast territory (e.g., more than 9000 bound nuclei in the nuclear landscape).  

Various mass models of nuclei, including macroscopic, macroscopic-microscopic and microscopic methods, have been successfully constructed and developed for local or global calculations~\cite{PhysRevLett.108.052501,MOLLER1995185,WANG2014215,PhysRevC.93.034337,PhysRevLett.102.152503,10.1143/PTP.113.785,XIA20181}. Nevertheless, the accurancies predicted by the majority of such theoretical models have yet to reach the desired level and lack the agreement, especially for the nuclei far from the stability line. How to obtain the satisfied mass information of nuclei in the extreme region is still an open problem, involving different physics modeling. 

Thanks for the rapid development of the computer technology and artificial intelligence techniques, the data-driven modeling based on machine learning has gained considerable attention and popularity in the last decade~\cite{RevModPhys.91.045002,RevModPhys.94.031003}. Indeed, this allows us to improve the model-prediction problem from another perspective. It is well known that the traditional physics-models usually contain a mathematical equation, which reflects the corresponding physical laws, and a set of model parameters/coefficients. At this moment, the role of experimental data is primarily used for fitting the model parameters, while it is directly able to determine the model itself in the data-driven modeling. Moreover, the data-driven approach can provide a more accurate prediction and its enormous potential has been noticed in many branches of science by the researchers~\cite{Silver2016,PhysRevD.97.056009}. In the early 1990s, machine learning algorithms have already been utilized for studying nuclear masses (e.g., see Refs.\cite{GERNOTH19931,ATHANASSOPOULOS2004222, doi:10.1142/S0217979206036053} and reference therein). There are still many successful applications in recent years~\cite{Ming2022,PhysRevC.105.064306,LE2023122707,PhysRevC.109.034318,NIU201848,PhysRevC.106.L021303,LI2024138385}. For instance, many algorithms, such as the Bayesian neural network~\cite{NIU201848}, the Levenberg--Marquardt neural network~\cite{Zhang_2017}, Gaussian processes\cite{PhysRevC.109.064322}, the decision tree algorithm\cite{Carnini_2020}, and the multilayer perceptron algorithm \cite{Y_ksel_2021}, have been successfully applied to refine nuclear mass models. It should be noted, however, that our present purpose is not to provide a detailed review on this subject but the interested readers can consult review articles, cf. e.g., Refs.~\cite{RevModPhys.94.031003,He2023}.    

In this project, we will probe the refined performance, e.g., the metrics of the rms derivation and its uncertainty, of the CatBoost machine-learning algorithm to the HFB mass models with different Skyrme parameters. With the advances in computing and storage, the large-scale calculations based on, e.g., the self-consistent HFB approach, have become practicable and several global mass tables~\cite{PhysRevLett.102.152503} have been provided by using the effective Skyrme forces, which, as effective interactions, are usually used within the fully microscopic self-consistent mean-field calculation (owing to its analytical simplicity). Typically, the rms deviations of these mass tables compared to the available experimental data are quite high and range between 2.0 and 5.0 MeV, depending on the interaction used in the calculations. Experimentally, the latest dataset of nuclear masses has been provided by the 2020 Atomic Mass Evaluation (AME2020)~\cite{Wang_2021}, which contains a total of more than 3500 nuclides but about 2500 nuclei with measured binding energies (the remaining ones have their binding energy estimated). The CatBoost algorithm, developed by Prokhorenkova et al.\cite{NEURIPS2018_14491b75}, replaces the traditional gradient estimation method with an order boosting, which reduced gradient bias and prediction shift, improved the accuracy and generalization ability of the model while reducing the happening chance of overfitting\cite{ZHANG2020125087}. So far, there are few works evaluating the different combinations between the HFB mass model with different Skyrme parameters and the CatBoost algorithm though numerous investigations on the evaluation of refined nuclear mass models by various machine learning algorithms have been made, e.g., see Refs.~\cite{Gao2021,PhysRevC.93.014311,PhysRevC.104.014315}. It will be somewhat interesting to explore whether the CatBoost algorithm can exhibt good performance on predicting nuclear mass. Presently, it should be stressed that what our primary concern is to evaluate the predictive power of such different combinations, focusing on their abilities of picking up the missing physics, rather than to obtain something else, e.g., the smallest rms deviation. 

{
Of course, it has hitherto been well known that there are some inevitable disadvantages in data-driven approaches. For instance, it is difficult for them to predict new or unknown physics and to construct physically interpretable models. Nevertheless, the benefits of the machine learning models outweigh these drawbacks to a large exent---they can combine physics that they learn in novel ways that is difficult to produce through standard physics modeling and give the predictions with similar even better accuracies compared to state-of-the-art theoretical models~\cite{RevModPhys.94.031003,Liu2025} and the uncertainty estimations which are widely stressed in theory~\cite{Zhang2021,Liu2017,Dobaczewski2014}.  Indeed, gaining insight into the underlying physics is still rather challenging from machine learning algorithms which often lack interpretability,  being regarded as “black-box” models. So far, it is powerful for machine learning models to learn the correlations between data features and target variables (labels), differring from physically based models which, focusing on the causal relations, usually capture the main representations of the physical processes. That is, moving from correlation to causation, equivalently, understanding the physics mechanism from machine learning, is still difficult. In principle, based on such correlational pattern recognition, it is somewhat insufficient for reliable decision-making and robust predictions in machine learning and, therefore, some new approaches to machine learning based on principles of causal reasoning are hope to provide a promising path forward in the future~\cite{Luo2020}. In the present project, it is supposed that the residual between theory and experiment is primarily related to model deficiency (associated to, e.g., the missing high-order interactions and/or many-body terms during the theoretical modeling, and the strength bias of model parameters from the fittings with different data constraints), and model uncertainty, originating from the error propagation of model parameters (note that the parameter error, different from the bias, refers to the uncertainty propagated from the uncertainty of experimental data in the fitting process). Generally, the former to a large extent represents the ``missing physics'' and could be picked up by mechine learning, while the later, usually exhibiting a random distribution, cannot be eliminated.
}


This paper is organized as follows. In Sec. \ref{sec:Theoretical_method}, we will briefly
introduce the theoretical methods, such as the bare HFB calculation and CatBoost algorithm. Sec.~\ref{sec:Results and Discussion} will be devoted to the main results and discussion. The predictive power of the bare and CatBoost-refined HFB models with different Skyrme forces will be presented. The refined performance of the CatBoost algorithm to the HFB models will also be evaluated and discussed, including the generalization ability to the 21 newly measured nuclei recently. Finally, the summary and concluding remarks will be provided in Sec.~\ref{sec:summary}.

\section{Theoretical method}
\label{sec:Theoretical_method}

In this section, we present a simple overview of the nuclear mass dataset, the HFB method, Skyrme parameter sets, and the CatBoost algorithms employed in this project. Details and different realizations of the method used can be found in Refs.\cite{STOITSOV200543,Erler2012,STOITSOV20131592} and references quoted therein.

\subsection{Nuclear mass dataset and the Skyrme HFB method}

The residuals, as the target variables in the trainning process of machine learning, depend on the hybrid data in both experiment and theory. Experimental mass dataset is provided by the 2020 Atomic Mass Evaluation (AME2020)~\cite{Wang_2021}, which comprises a total of more than 3500 nuclides but about 2500 nuclei have their binding energy measured (the remaining ones have the estimated values). In the present study, we define the experimental dataset by using those nuclei who have the measured binding energy. Strictly speaking, there are 2457 nuclei with $Z, N \geq 8$ in total for this dataset. Further, the information in this set is rather precise with an average reported mass uncertainty of roughly 25 keV. Theoretically, for simplicity, we take the calculated mass tables from Ref.~\cite{dfttable}  instead of repeating the large-scale calculations made using the Hartree-Fock-Bogoliubov solver HFBTHO~\cite{STOITSOV200543,Erler2012}. The different mass tables are labeled by the adopted Skyrme parameter sets including SkM*~\cite{BARTEL198279}, SkP~\cite{DOBACZEWSKI1984103}, SLy4~\cite{CHABANAT1998231}, SV-min~\cite{PhysRevC.79.034310}, UNEDF0~\cite{PhysRevC.82.024313}, and UNEDF1~\cite{PhysRevC.85.024304}. In priciple, the information in these mass tables is less precise because of the approximations made in the modeling of atomic nuclei. However, these datasets can provide the valuable new sources of information for nuclei that have not yet been measured, even for the extreme regions (e.g., far from the stability line or superhavey). 

The Skyrme HFB methods are standard to date and we would like to briefly introduce them and provide the necessary references instead of the detailed review here. In these HFB mass calculations, the quasi-local Skyrme functions was used in the particle-hole channel and the density-dependent zero-range pairing term was used in the pairing channel. The self-consistent HFB equation of the nucler density functional theory (DFT) were solved with the code HFBTHO~\cite{STOITSOV200543} as further optimized in Refs.~\cite{PhysRevC.82.024313,PhysRevC.85.024304}. It is well known that the nuclear DFT based on the self-consistent mean-field approach is an important microscopic tool~\cite{Stoitsov2003,Stoitsov2009} and the adopted mass data were calculated within such a DFT framework. The present code solves the nonlinear HFB equations in configuration space by expanding self-consistent eigenstates in a large basis of the deformed harmonic oscillator. The axial symmetry and parity are imposed to reduce the dimension and complexity of the nuclear mean-field problem, neglecting the effect of triaxial and reflection-symmetric ground-state deformations (which is expected to be minor). The harmonic oscillator states originating from the 20 major oscillator shells are used for constructing the single-particle basis. The calculated results are already stable enough with respect to a possible enlargement of this basis space. To restore approximately the particle number symmetry broken in the HFB method, the variant of the Lipkin-Nogami scheme are used during the calculations~\cite{Stoitsov2003}.


Concerning the Skyrme interactions, they were originally designed for describing the self-consistent HF mean field and their parameters are generally determined by adjusting selected bulk properties of infinite matter and ground states of finite nuclei. This effective interactions between nucleons can be represented by the energy density function, which depends on total (neutron-plus-proton) and isovector (neutron-minus-proton) densities and currents. With a short-range pairing term, the Skyrme energy functional is usually parameterized by up to 14 coupling constants. For predicting the whole nuclear chart, six sets of Skyrme parameter sets as described above are adopted for the HFB mass calculations. Without entering into details addressing how to extract the Skyrme parameter sets, let us limit ourselves to merely reminding the interested readers about their primary consideration. For instance, focusing on surface energy and fission barriers, it is developed for SkM*\cite{BARTEL198279}; for the purpose of a simultaneous description for the mean and pairing fields, SkP parameter set is developed~\cite{DOBACZEWSKI1984103}; with a bias on neutron-rich nuclei and properties of neutron matter, SLy4 is optimized~\cite{CHABANAT1998231}; SV-min\cite{PhysRevC.79.034310} is adjusted according to a variety of data on spherical nuclei (inculding diffraction radii and surface thickness); considering data on spherical and deformed nuclei, UNEDF0 is developed~\cite{PhysRevC.82.024313}; and the data set of UNEDF1\cite{PhysRevC.85.024304} also takes the excitation energies of fission isomers into account. Some more details with regard to adaptation to the present illustrations can be found in Refs.\cite{dfttable,Erler2012} and references therein.  

\subsection{The CatBoost algorithm}
\label{subsec:CatBoost_features}

The CatBoost algorithm~\cite{dorogush2018,NEURIPS2018_14491b75} based on decision tree is an ensemble method and used for creating new model to reproduce the mass residuals and predict yet to be measured masses with sound uncertainties. This algorithm is a novel variant of the gradient-boosting decision tree (GBDT) method developed by Prokhorenkova et al. in 2018\cite{NEURIPS2018_14491b75}. As it is known that the decision tree likes a tree with a root node, internal nodes, and leaf nodes. It generally uses simple rules to start from the root node and branch out, going through internal nodes, to finally end up in the leaves. The ensemble learning technique will use a sequence of decision trees and each tree will learn from the previous one and affect the next one to improve the model and build a strong learner. As an innovative algorithm for processing categorical features, CatBoost is an improved implementation under the GBDT framework~\cite{dorogush2018,NEURIPS2018_14491b75}. At present, this method has already become the popular and powerful technique with outstanding performance for dealing with not only classification but also regression problems by variables with continuous and discrete values, respectively. In this method, the model accuracy, generalization ability and robustness can usually be improved since the gradient bias and prediction shift problems are dealt with efficiently (cf. e.g., Ref.~\cite{NEURIPS2018_14491b75} for an in-depth analysis of this algorithm in terms of classification and regression for different tasks).

The main idea of the CatBoost algorithm is that a strong regressor can be created by iteratively combining weak regressors\cite{Ding2021ACA}. Indeed, by using an ensemble learning strategy, it can help to combine weaker regression models to form a robust regression model. CatBoost has strong learning capabilities to deal with highly nonlinear data~\cite{DENG2020106344,ABBASNIYA2022108382}. 
During the training process, for instance, the loss function $L(y,F(\mathbf{x}))$ can be minimized by gradient descent:
\begin{align}
	h^{ t } = 
	\underset{ h \in H }{ \arg \min } \mathbb{ E }
	\left\{ L \left( y , F^{ t - 1 } ( \mathbf{ x } ) 
	                 + h ( \mathbf{ x } ) \right) \right\}          \label{Eq.1}
\end{align}
where $y$ denotes the output and $h$ is a gradient step function selected from $H$, i.e., a family of functions. The step function can be calculated as follows:
\begin{align}
	h ( \mathbf{ x } ) 
	= \sum_{ j = 1 }^{ J } b_{ j } 
	  \mathbbm{ 1 }_{ \left\{ \mathbf{ x } \in R_{ j } \right\} }
	                                                                \label{Eq.2}
\end{align}
where $R_j$ denotes the disjoint regions that correspond to the tree's leaves. $b_j$ is the predictive value of the region, and $\mathbbm{1}$ is an indicator function, e.g., cf Ref.\cite{NGUYEN2022114768} for more details. For performing the training procedure, the dataset is randomly split into a training and testing sets. The CatBoost puts the training set into the algorithm for learning and, then, randomly arranges all these data and filters out samples with the same category from all features. When numerically transforming the characteristics of each sample, the target value of the sample is first calculated before the sample, and the corresponding weight and priority are added. Once the CatBoost modeling has the highest performance on the training set, we use such a trained model on the test set.
In addition, it should be noted that the expectation in Eq.(\ref{Eq.1}) is approximated by the same dataset. This may lead to prediction shift and gradient bias and CatBoost can resolve these issues by employing the ordered boosting method, improving the model robustness and generalization ability~\cite{DENG2020106344,NGUYEN2022114768}.
 

\section{Results and Discussion}
\label{sec:Results and Discussion}

\begin{table}[htbp]
	\caption{Selection of characteristic quantities.}
	\begin{ruledtabular}
		\renewcommand
		\arraystretch {1.5}
		\centering
		\begin{tabular}{p{3cm}p{5cm}}
			Feature&Description\\
			\specialrule{0.05em}{1pt}{1pt}
			$Z$&Proton number\\
			$N$&Neutron number\\
			$N/Z$&Ratio of neutrons to protons\\
			$Z_{eo}$&$Z_{eo}$ is 0 if Z is even and 1 if Z is odd\\
			$N_{eo}$&$N_{eo}$ is 0 if N is even and 1 if N is odd\\
			$|Z-m|$&Distance between proton number and the nearest magic number;

			$m\in\{8,20,28,50,82,126\}$\\
			$|N-m|$&Distance between neutron number and the nearest magic 
			number; 

			$m\in\{8,20,28,50,82,126,184\}$\\
		\end{tabular}
	\end{ruledtabular}                                     \label{table:TABLE-1}
\end{table}

\begin{figure}[htbp]
\centering
\includegraphics[width=1.0\linewidth]{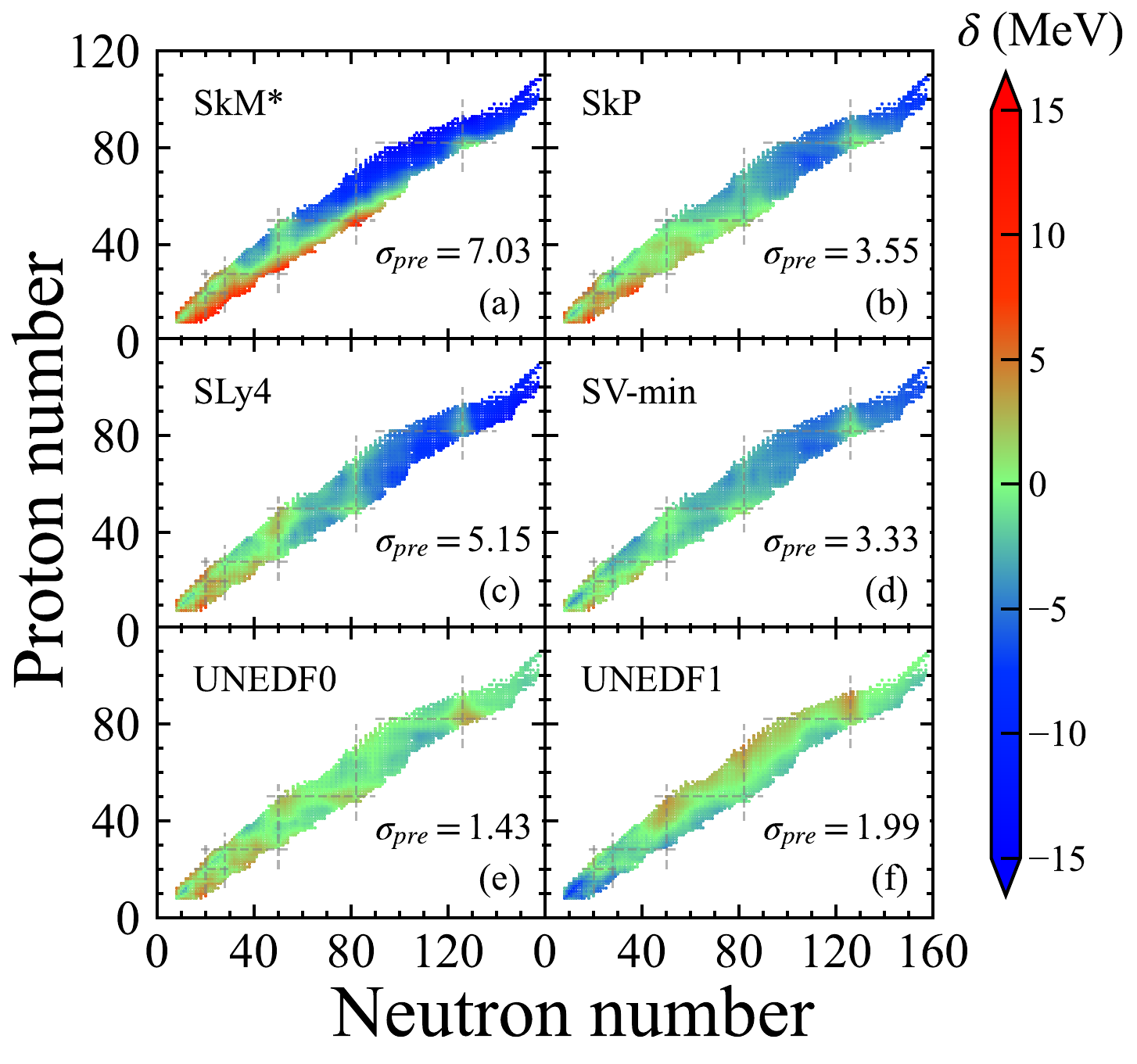}	
\caption{(Color online) Residual $\delta(Z,A)=B_{\mathrm{HFB}}-B_{\mathrm{exp}}$ predicted by HFB calculations with different Skyrme parameters [e.g., (a) SkM*\cite{BARTEL198279}, (b) SkP\cite{DOBACZEWSKI1984103}, (c) SLy4\cite{CHABANAT1998231}, 		(d)SV-min\cite{PhysRevC.79.034310}, (e) UNEDF0\cite{PhysRevC.82.024313} and (f) UNEDF1\cite{PhysRevC.85.024304}]. The experimental binding energy $B_{\mathrm{exp}}$ are taken from AME2020\cite{Wang_2021}. As seen in each subfigure, the rms deviation $\sigma_{pre}$ is also shown. }
															      \label{Fig.1}
\centering
\includegraphics[width=1.0\linewidth]{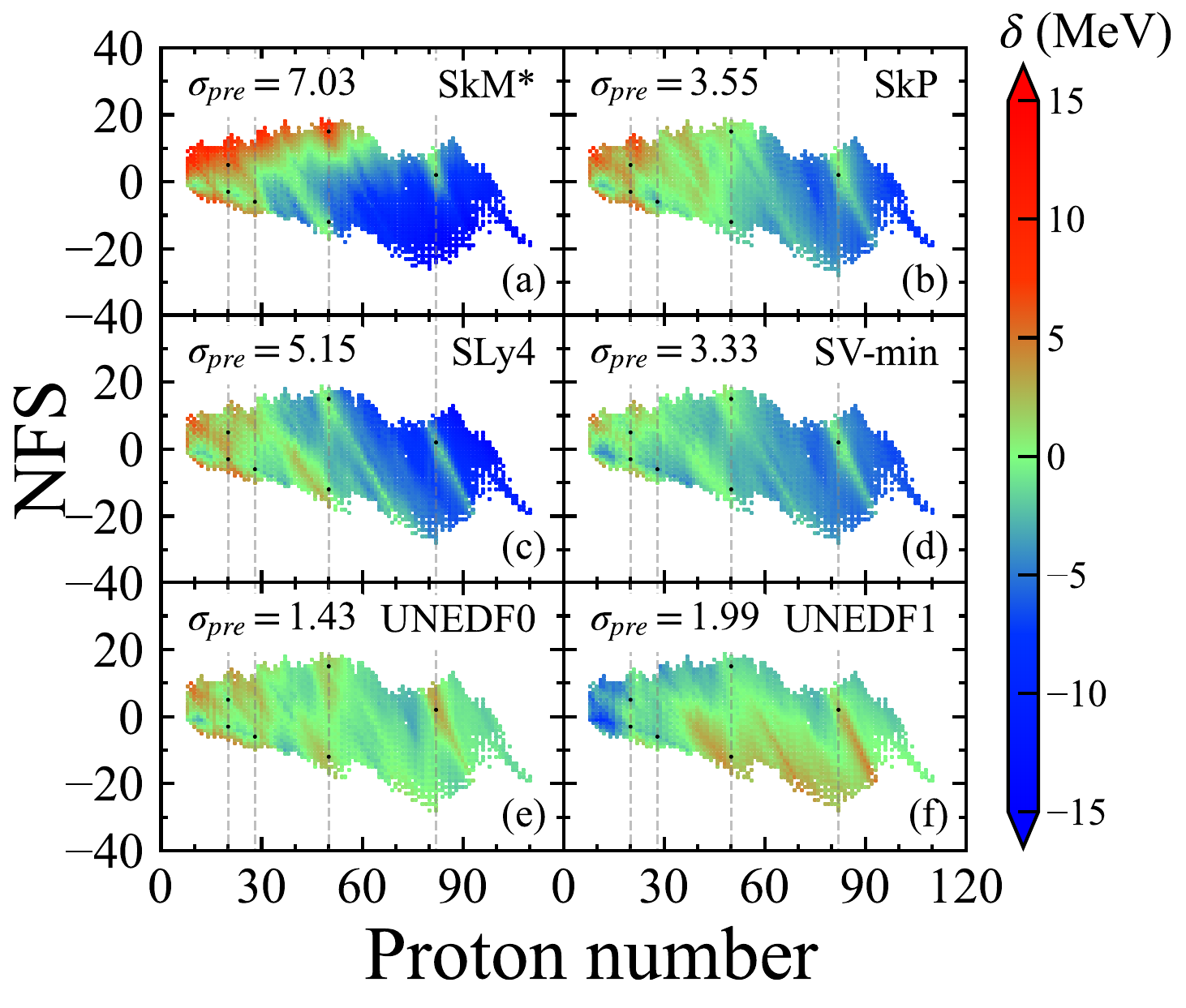}
\caption{(Color online) Similar to Fig.~\ref{Fig.1} but $Z$ vs NFS plane. The dash lines illustrate the positions of the proton magic numbers (e.g., 20, 28, 50, and 82). See text for more details.}
	                                                              \label{Fig.2}
\end{figure}

Let us begin by introducing the selected features in the present project. As is well known, the selection of appropriate features for the data-driven, e.g., machine-learning model is extremely important since they, as inputs during training, can significantly impact the model performance and directly related to the model interpretability. In the present work, considering the available literature\cite{Gao2021,PhysRevC.106.L021301, LI2024138385,Ming2022,LE2023122707,NIU201848, PhysRevC.109.064322,PhysRevC.105.064306,Carnini_2020}, we try to incorporate those relevant nuclear features that may influence mass predictions. In the present project, seven features are contained, as one can see in Table~\ref{table:TABLE-1}. Note that the bulk properties are defined as the proton (neutron) number $Z$ ($N$). The term $N/Z$ is the ratio of neutrons to protons. The term $|Z-m|$ ($|N-m|$), which measures the distance between proton (neutron) number and the nearest magic number, is related to the shell structure. Here, the proton and neutron magic numbers are taken as $m$ = 8, 20, 28, 50, 82, 126, 184 (only for neutrons). Two other features $Z_{eo}$ and $N_{eo}$ reflect the pairing information, where $Z_{eo} (N_{eo})$ is 0 if $Z (N)$ is even and 1 if $Z (N )$ is odd\cite{PhysRevC.106.014305}.

Following the procedure, e.g., in Ref.~\cite{Gao2021}, the residual $\delta (Z,N)$ [$\equiv B_{th}(Z, N) - B_{exp}(Z, N)$] between the theoretical prediction and the experimental binding energy, as the target variable, is used for training the machine-learning model. Once the behavior of the residual $\delta (Z,N)$ is captured, the binding energy of the nucleus  outside the training set can be calculated using the relationship $B_{ml}=B_{th}(Z, N)-\delta (Z,N)$. In general, the rms deviation $\sigma_{rms}$ ($\equiv \sqrt{\frac{1}{n}\sum_{i=1}^n(B^i_{th}-B^i_{exp})^2 }$) of the theoretical model can be significantly improved after the machine-learning refinement (e.g., replacing $B_{th}$ by $B_{ml}$). For comparison, together with the rms deviations, Fig.~\ref{Fig.1} presents the residual $\delta (Z,N)$ between the HFB calculations with SkM*\cite{BARTEL198279}, SkP\cite{DOBACZEWSKI1984103}, SLy4\cite{CHABANAT1998231}, SV-min\cite{PhysRevC.79.034310}, UNEDF0\cite{PhysRevC.82.024313} and UNEDF1\cite{PhysRevC.85.024304} Skyrme parameters, and the experimental data measured in AME2020\cite{Wang_2021} with $Z, N \geq 8$. One can notice that the used Skyrme parameter sets reproduce total binding energies with the rms deviation of the order of 1-7 MeV, agreeing with the report in Ref.~\cite{Erler2012}. Moreover, from south-east to north-west, there is a decreasing trend for the residuals for each subfigure. Similarily, Fig.~\ref{Fig.2} exhibits the same residuals in the $Z-NFS$ plane, where $NFS$ means the Neutron number From the beta Stability line (e.g., the Green formula is adopted
~\cite{AESGreen}). From a new perspective, application of such coordinate transformations may help the display beter. The residual distributed properties can be seen from these two figures, more clearly from the later.

\begin{table}[htbp]
\footnotesize
\caption{Eight arbitrarily selected feature spaces adopted for model training. More explanations, see the text.}
\renewcommand
\arraystretch {1.5}
\centering
\begin{tabular}{cc}
\hline
\hline
\textbf{Model}&\textbf{Feature Space}\\
\hline
M$_1$ &$Z, N$\\
M$_2$ &$Z, N, Z_{eo}, N_{eo}$\\
M$_3$ &$Z, N, |Z-m|, |N-m|$\\
M$_4$ &$Z, N, Z_{eo}, N_{eo}, |Z-m|, |N-m|$\\
M$_5$ &$Z, N, N/Z$\\
M$_6$ &$Z, N, N/Z, Z_{eo}, N_{eo}$\\
M$_7$ &$Z, N, N/Z, |Z-m|, |N-m|$\\
M$_8$ &$Z, N, N/Z, Z_{eo}, N_{eo}, |Z-m|, |N-m|$\\
\hline
\hline
\end{tabular}                                          \label{table:TABLE-2}
\end{table}

\begin{figure}[htbp]
\centering
\includegraphics[width=1.0\linewidth]{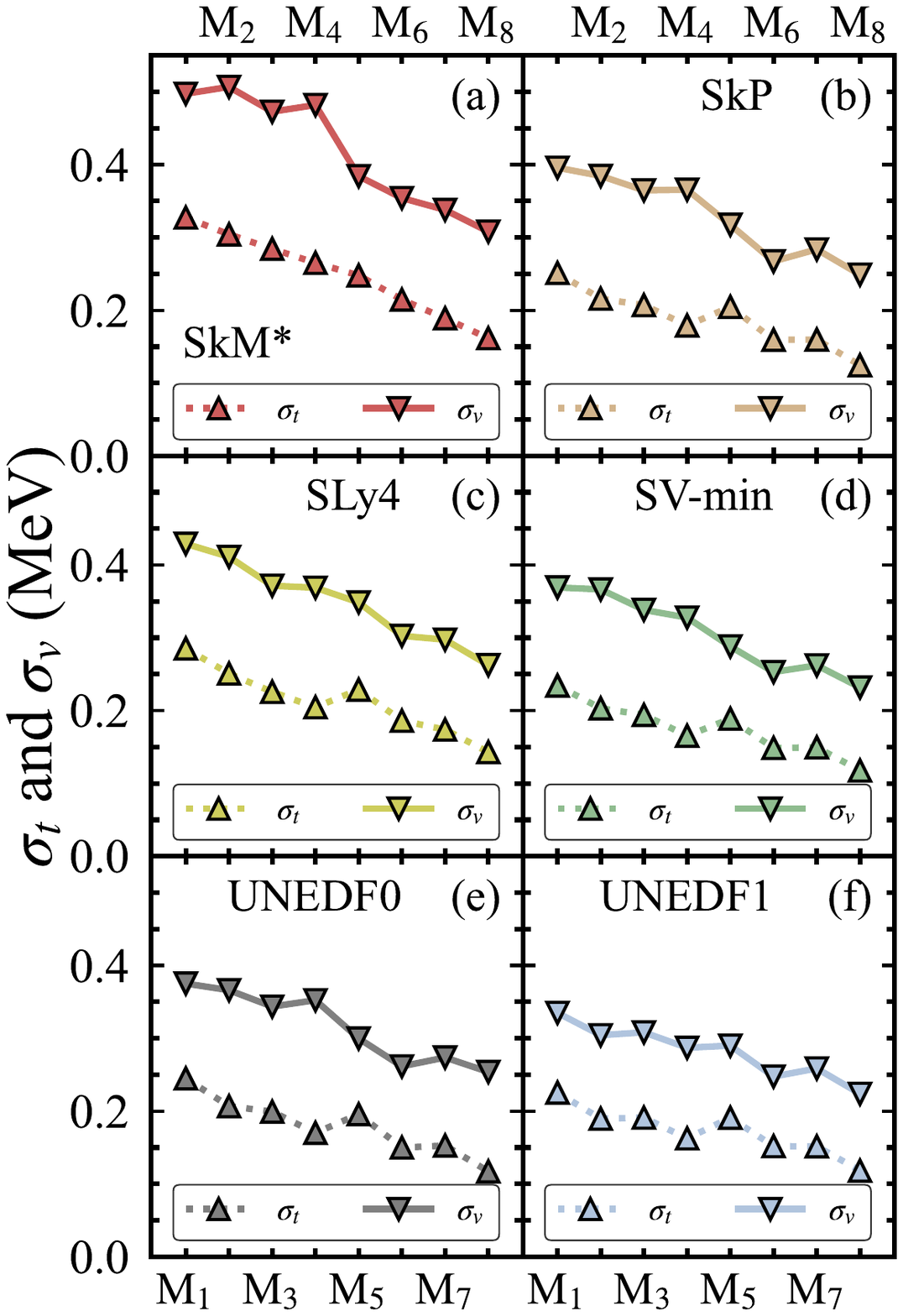}
\caption{(Color online) With different Skyrme parameters, the calculated rms deviations of the training set ($\sigma_t$) and validation set ($\sigma_v$) plotted as a function of the selected feature groups shown in table~\ref{table:TABLE-2}.}
															       \label{Fig.3}
\end{figure}

\begin{figure}[htbp]
\centering
\includegraphics[width=1.0\linewidth]{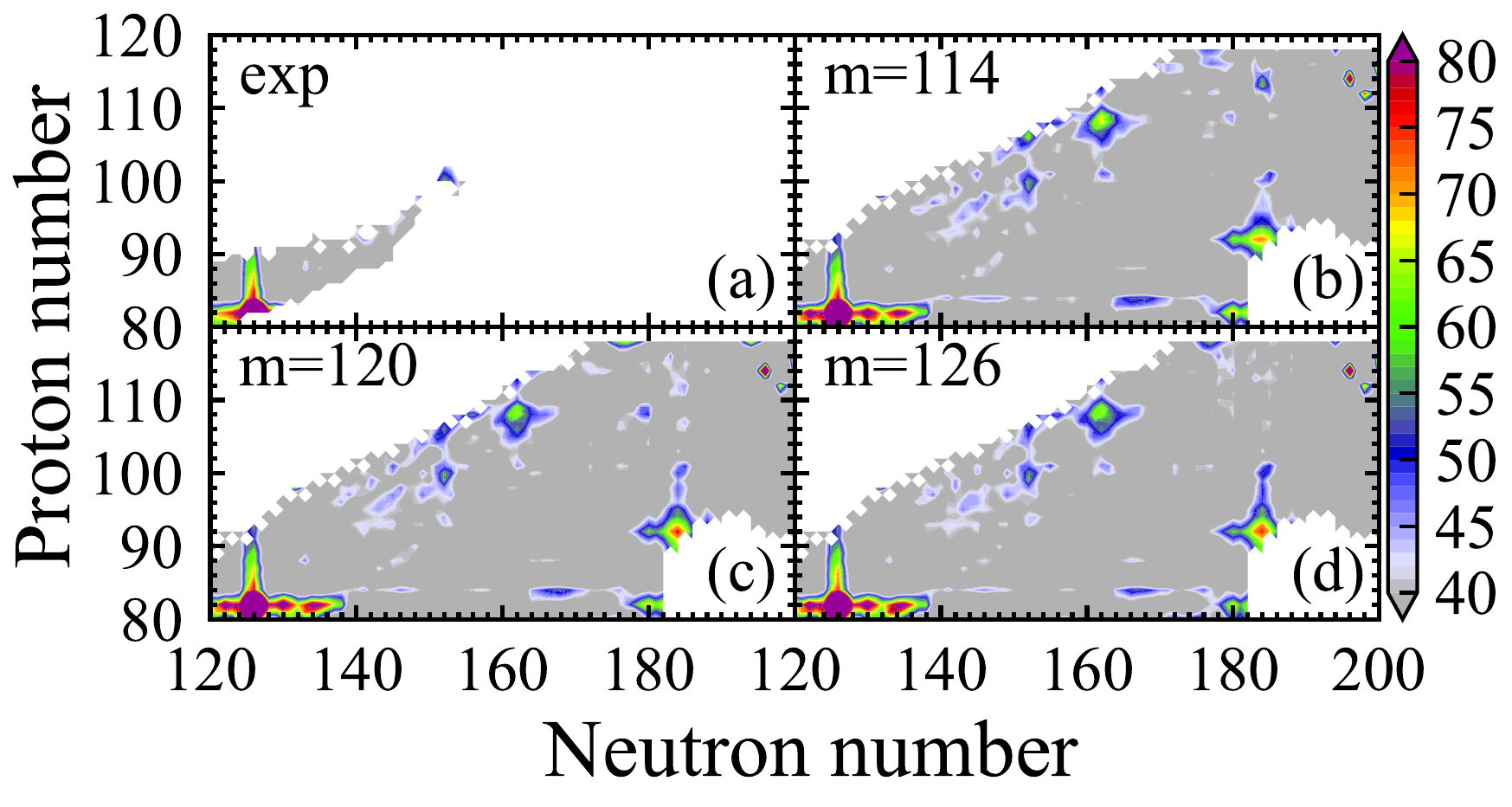}
\caption{(Color online) Contour plots of shell gaps in nuclei scaled by $A^{1/2}$ from the experimental data (a) and CatBoost-refined HFB masses with the UNEDF1 Skyrme parameters by using proton magic numbers 114 (b), 120 (c) and 126 (d) beyond $Z= 82$. }
	                                                               \label{Fig.4}
\end{figure}

To test the effect of the different combinations (as seen in Table~\ref{table:TABLE-2}) of the input features in meachine learning and obtain a more robust estimate of the model’s performance, e.g., the metrics $\sigma_{rms}$, we use the cross-validation technique, e.g., $k$-fold cross validation ($k=10$ in the present case), in this study. Note that the more folds one uses in $k$-fold cross-validation the lower the bias of the mean-squred error but the higher the variance; on contrary, the fewer folds one uses the higher the bias but the lower the variance. For simplicity, we randomly divide the entire dataset into the training and testing sets at a ratio 4:1 during the test. Further, the training set is devided ten folds (namely, ten subsamples) and then, using default model hyperparameters, train the model on the nine folds (2211 nuclei, 90\%) but validate the model on the remaining one fold (246 nuclei, 10\%). The eight input-combinations labeled by $M_1$ to $M_8$ are tested and the results are presented in Fig.~\ref{Fig.3}. One can see that all the rms deviations calculated by the refind mass models are significently improved. Even, just considering the proton and neutron numbers as feature variables, the rms deviations for six sets of Skyrme parameters can reduce to less than 0.5 MeV in the testing sets. It seems that, taking all the features as inputs (namely, the combination $M_8$) in the Catboost algorithm, we can obtain the smallest rms deviations (around 0.2-0.3 MeV) for each parameter set. {In addition, as seen in Table~\ref{table:TABLE-1}, the proton magic number 126 is adopted beyond 82 but there is no general agreement between different models. For instance, the shell closures at $Z = 114$, 120 and 126 are respectively obtained using the macro–micro Woods–Saxon (WS) calculations~\cite{Sobiczewski1966}, the Skyrme (or Gogny) Hartree–Fock–Bogoliubov (HFB) methods~\cite{Bender1999,Sobiczewski2007}, and the relativistic mean-field (RMF) methods~\cite{Sobiczewski2007}. To see the influence of selecting different proton magic numbers (e.g., 114, 120, or 126) in the region of superheavy nuclei on the mass predictions, similar to in Ref.~\cite{Mo2014} (cf. Fig.~9 therein), for heavy nuclei, Fig.~\ref{Fig.4} illustrates the contour plots of scaled shell-gap $\Delta (N,Z) A^{1/2}$ calculated by the experimental data and the CatBoost-refined HFB masses with the arbitrarily selected UNEDF1 parameters. Since the smooth macroscopic part in the nuclear binding energy is cancelled out through the mass difference, the shell gap as a sensitive probe could be used for investigating the fine structure of nuclei caused by the residual shell effects. From Fig.~\ref{Fig.4}, one can see, using different proton magic numbers in the superheavy region, that the scaled shell-gap values are significantly larger for nuclei with well-known magic numbers (e.g., at $Z =$ 82 and $N =$ 126, 184) than those of open-shell nuclei due to the shell effects. Though the superheavy nucleus $^{298}_{114}$Fl$_{184}$ seems to exhibit the slightly prominent shell gap in Fig.~\ref{Fig.4}(b), there is no evident difference between these countour plots which may, as excepted, originate from the fact that different proton magic numbers just affect the features of fewer nuclei in the training set. Therefore, for comparing with some of the results, e.g., in Ref.~\cite{Gao2021}, we continue to use the $M_8$ combination and proton magic number 126 beyond 82 as inputs to further conduct the following machine-learning modeling.}

\begin{figure}
\centering
\includegraphics[width=1.0\linewidth]{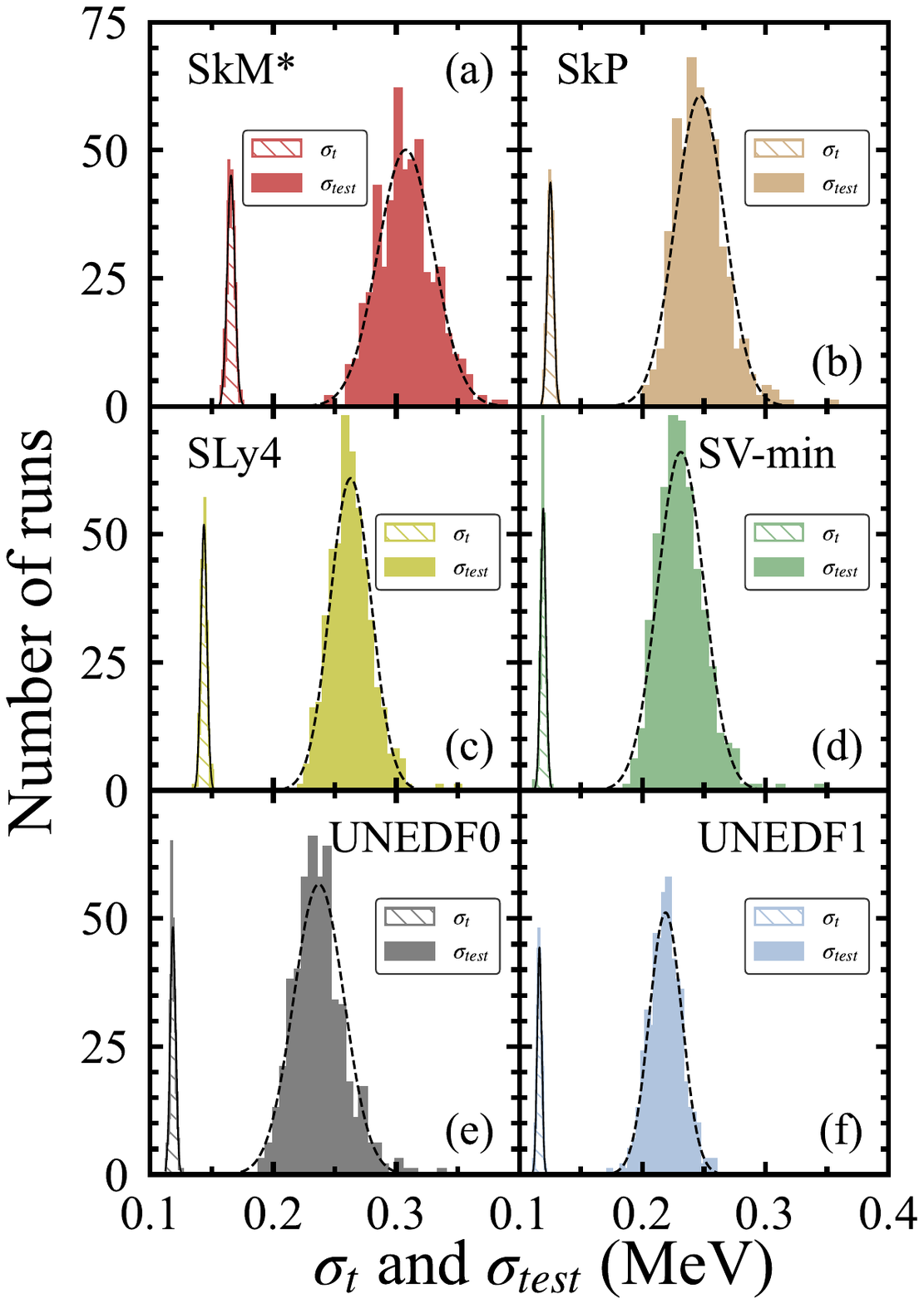}
\caption{(Color online) Distribution of the rms deviations ($\sigma_t$ and $\sigma_{test}$ for training and testing sets, respectively). Data are from 500 runs for each Skyrme parameter set (SkM*, SkP, SLy4, SV-min, UNEDF0 and UNEDF1). For each run, the entire dataset (2457 nuclei) was randomly split into training and testing sets with a ratio $4:1$. The CatBoost model was then trained with the default hyper-parameters.
	}
																 \label{Fig.5}
\centering
\includegraphics[width=1.0\linewidth]{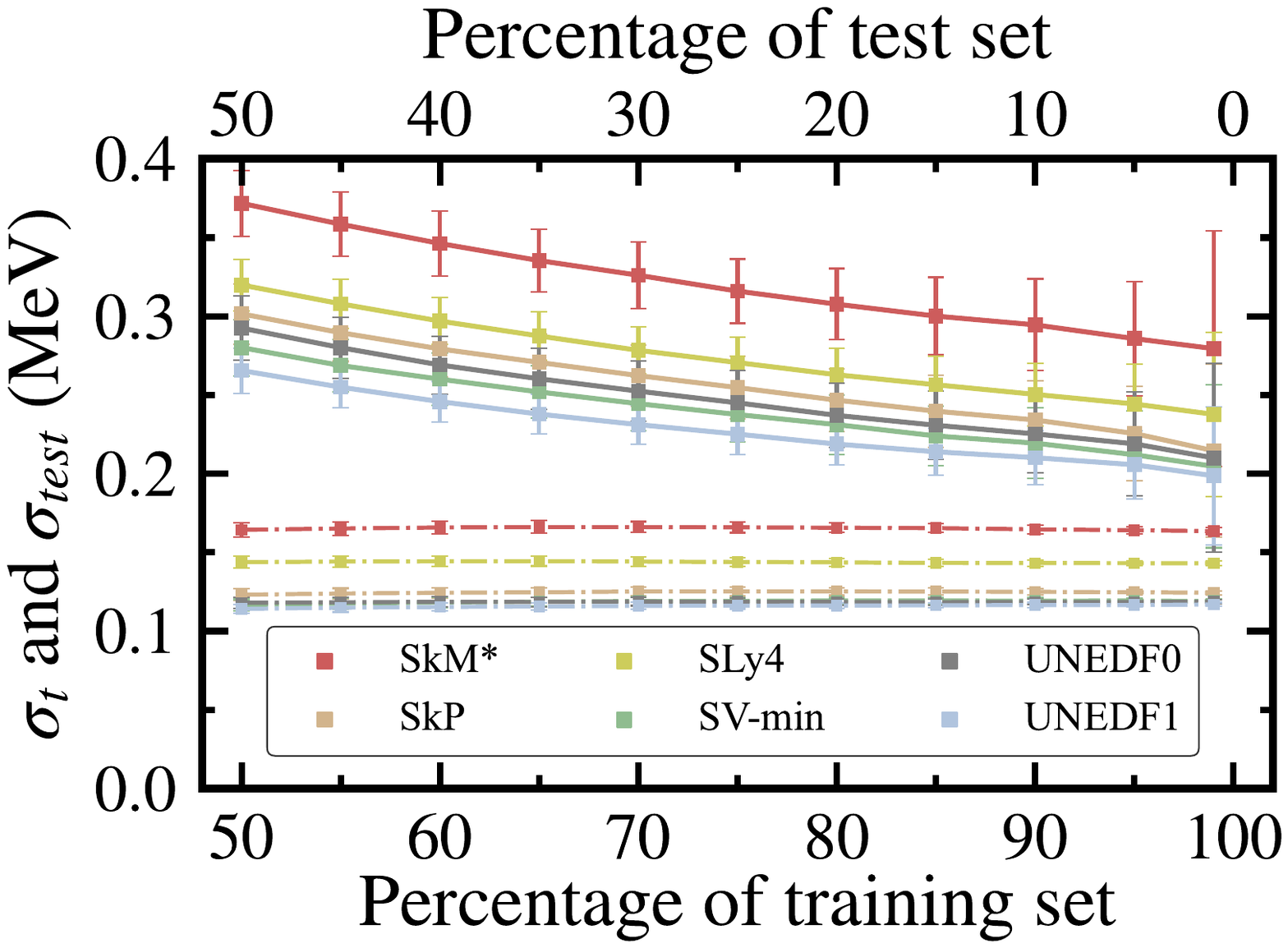}
\caption{(Color online) The rms deviations for testing sets ($\sigma_{test}$, solid lines) and training sets ($\sigma_t$, dotted lines) plotted as a function of the size of the training/testing set. The test was performed with the default hyper-parameters.}
												                 \label{Fig.6} 
\end{figure}

To investigate the effect of the training set size on the predictive performance (e.g., the rms deviation and its uncertainty) on the testing set, Figs.~\ref{Fig.5} and \ref{Fig.6} present the calculated results with the default hyperparameters and the various sizes of training data for six mass tables labeled by the adopted parameter sets. It should be noted that different size of training data sets were created by randomly splitting the entire dataset according to a special ratio. For each Skyrme parameter set, as examples, Fig.~\ref{Fig.5} exhibits the distributions of the rms deviations for the training and testing data sets by using the data of 500 repetitions at the spliting ratio 4:1. That is, for each recepition, the entire data set is randomly split into a training set (80\%) and a testing set (20\%). Based on eleven training data sets, Fig.~\ref{Fig.6} illustrates the evolution of the rms deviations and their uncertainties with the different percentages of the training or testing sets. It can be seen from these two figures that the rms deviations exhibit the approximately Gaussian distributions for both the training set and the testing set. On the testing set, the relative error is less than 10\%, displaying a stable prediction. With the increasing size of the training set, the rms deviations corresponding to different Skyrme parameters have an overall decreasing trend, while their uncertainties increase slightly.  One can also notice that the typical value of rms deviations on the training data set is only $\sim0.04$-0.1 MeV, which is similar to Ref.\cite{Gao2021}. However, the $\sigma_t$ in this work is slightly smaller than that of LightGBM~\cite{Gao2021}. The gradual growth of error bar on the testing set, e.g., with its decreasing size, is primarily caused by the statistical uncertainty. As expected, the rms deviation on the training set gradually decreases with the increasing size. Agreeing with the illustrations in Figs.~\ref{Fig.3}, the calculated results further indicate that the selected input features exhibt the good performance for the machine-learning modeling. It also can be understand that, due to the size limitation of the overall dataset, the CatBoost model will be able to learn more informaiton by reducing the capacity of the testing set and the rms deviation will certainly be decreased. About the uncertainty of the rms deviation, as seen in Fig.~\ref{Fig.6}, it will increase with a reduce in the percentage of the testing set since the less the nuclei in the testing set, the larger the fluctuations of the rms deviations.

\begin{table}[htbp]
\caption{Selection of hyper-parameter tuning space for the CatBoost algorithm. Several frequently used hyper-parameters are considered, including the tree depth ($d$), the regularization coefficient $\Lambda_{reg}$, the learning rate $l_r$ and the iteration numbers $iterations$. See text for more explantions.}
	\renewcommand
	\arraystretch{1.5}
	\centering
	\begin{tabular}{lccccccc}
		\hline\hline
		Hyper-parameter & $d$ & &$\Lambda_{reg}$ && $l_r$ && $iterations$ \\
		\hline
	   Domain & $[3, 10]$ && [3, 9] && [0.02, 0.50]&& [1000, 5000] \\
	   Increment & $1$ && $3$ && $0.02$ && $2000$ \\
		\hline\hline
	\end{tabular}                                          \label{table3}
\end{table}
%

Hyper-parameters tuning is a critical step in training machine-learning models, which can significantly affect a machine-learning model's generalization performance, avoid overfitting and decrease model complexity. Though CatBoost-refined HFB models with default hyperparameter sets can give the rather high accurancy (cf. Fig.~\ref{Fig.3} or ~\ref{Fig.6}) for each Skyrme parameter set, it is expected that the suitable hyperparameter tuning will be able to further improve the model performance in the practical project. Certainly, the 10-fold cross-validation technique has been samely used for hyperparameter tuning during the model training of CatBoost. The average rms value across the 10 folds is used as the evaluation metrics for hyperparameters. Generally speaking, it is a time-consuming and complex task to find the optimal combination of hyperparameters, especially when dealing with a large range of hyperparameters and complex models. Up to now, though some hyperparameter optimization softwares, such as Hyperopt~\cite{Bergstra_2015} and Optuna~\cite{optuna_2019}, have been developed for trying to perform such kind of tasks, it is still necessary to exhibit some details related to the hyperparameter tuning. For instance, using a grid search approach, we scan over a given hyperparameter space, as illustrated in Table~\ref{table3}, including the learning rate $l_r$, the depth $d$ of the tree, the coefficient $\Lambda_{reg}$ for the $L2$ regularization term of the cost function and iteration times~\cite{NEURIPS2018_14491b75}. These hyperparameters are often supposed to be important for the CatBoost algorithm in the literature. For instance, the tree $\mathit{depth}$ will control model complexity at an appropriate level to achieve a suitable fitting by providing a good weak learner for the gradient boosting procedure~\cite{Friedman2001GreedyFA}. The $\mathit{learning\_rate}$ parameter governs the step size at which the model learns during the gradient boosting process. Its value is between 0 and 1 that determines how much each new tree in the ensemble contributes to the final prediction. A smaller learning rate (closer to 0) means that each new tree added to the ensemble has a smaller influence on the final prediction. On the other hand, a larger learning rate (closer to 1) allows each new tree to have a greater impact on the final prediction, leading to faster convergence. However, a learning rate that is too large can cause the model to overfit to the training data. 
                               
\begin{figure}[htbp]
\includegraphics[width=1.0\linewidth]{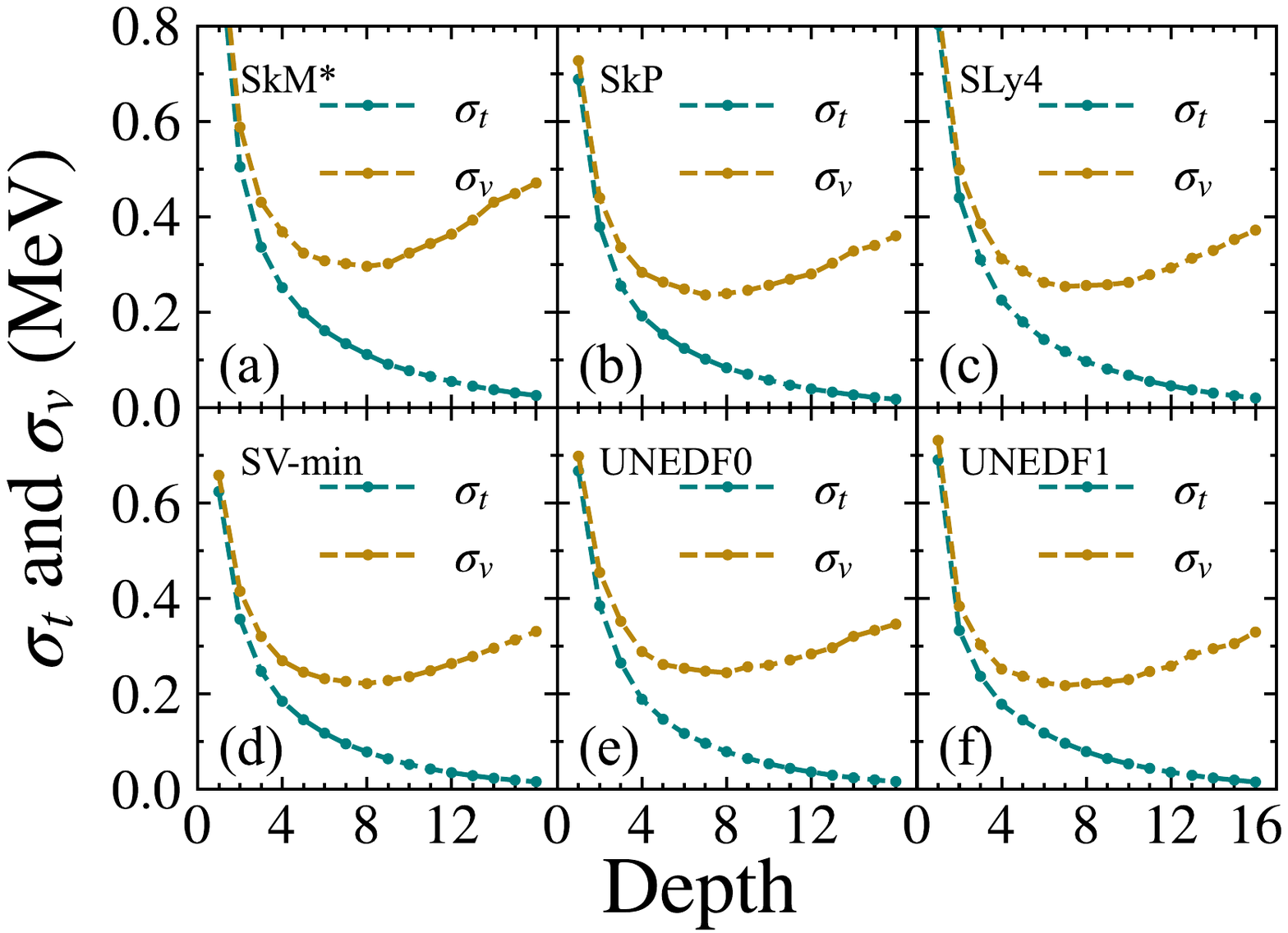}\\
\caption{(Color online) The rms deviations ($\sigma_t$, $\sigma_v$ for training and validation set, respectively) as functions of the $depth$ parameter. Other hyper-parameters are set to the default values. See text for more details.}
	                                                               \label{Fig.7}
%
\centering
\includegraphics[width=1.0\linewidth]{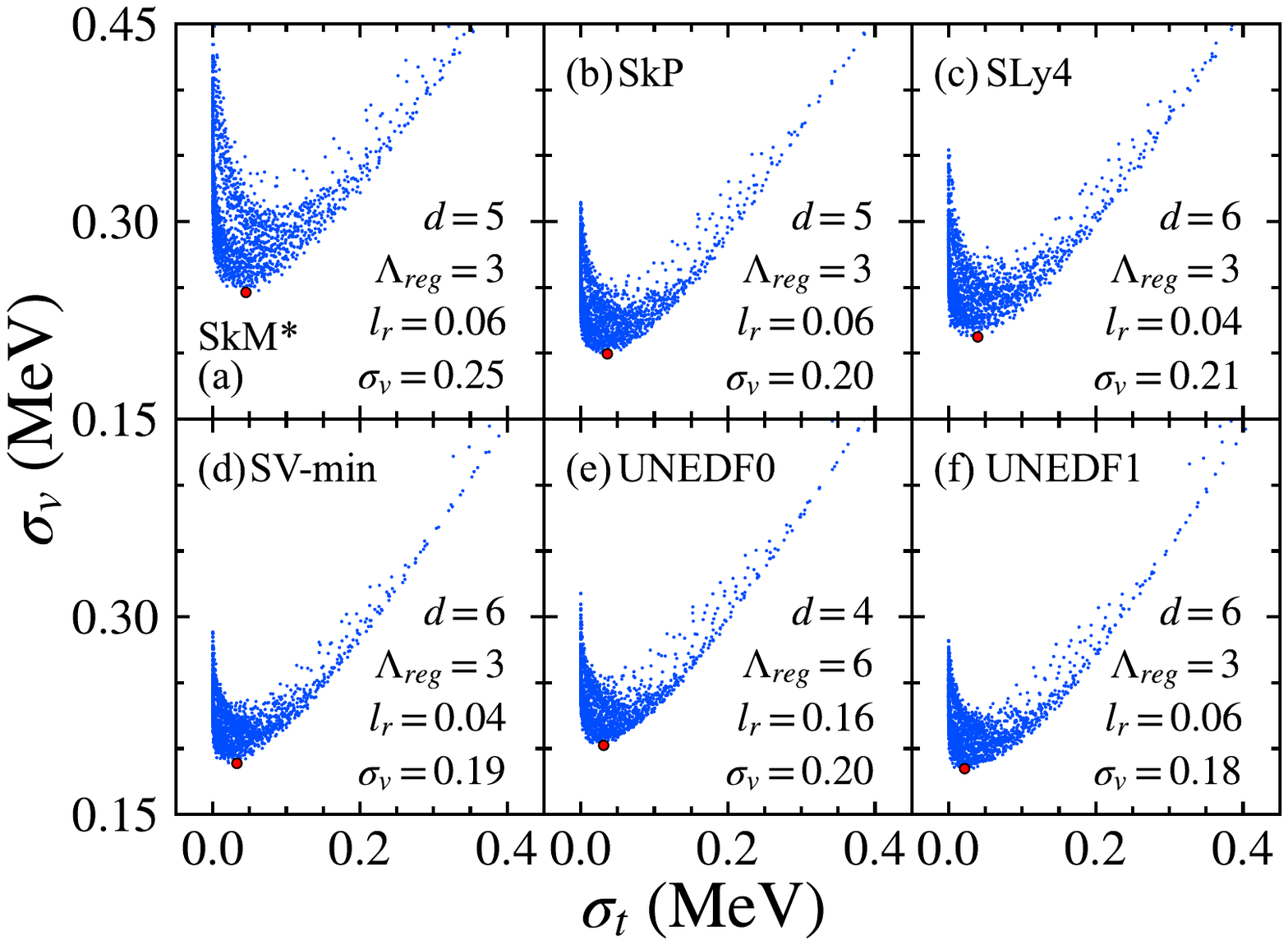}
\caption{(Color online) Scatter plots of the rms deviations in the $\sigma_t$ vs $\sigma_v$ plane. All the points, calculated in three dimensions ($d$, $\Lambda_{reg}$, $l_r$) of the selected hyper-parameter space, are displayed in this plot. The red dot in each subfigure indicates the minimum value of the rms deviations. The hyper-parameters calculated for the minimum correspond to the so-called optimal hyper-parameter values (together with the rms deviation $\sigma_v$, the obtained optimal hyper-parameters are listed in each subfigure). In addition, it should be noted that the iteration numbers $iterations$ for all optimal hyper-parameter groups are 5000.}
															       \label{Fig.8}
\end{figure}

\begin{table}
\caption{The optimal hyper-parameters for the CatBoost algorithm optimized by grid search and early stopping method during refining the HFB model with different Skyrme parameter sets.}
\begin{ruledtabular}
\renewcommand
\arraystretch {1.5}
\centering
\begin{tabular}{lcccc}
Parameter&$d$ & $l_r$ &$ \Lambda_{reg}$ &iterations\\
\hline 
SkM*    & 5 & 0.06 & 3 & 3916 \\
SkP     & 5 & 0.06 & 3 & 3078 \\
SLy4    & 6 & 0.04 & 3 & 4000 \\
SV-min  & 6 & 0.04 & 3 & 3834 \\
UNEDF0  & 4 & 0.16 & 6 & 1374 \\
UNEDF1  & 6 & 0.06 & 3 & 2496 \\
\vspace{-0.50cm}
\end{tabular}
\end{ruledtabular} 
	                                                           \label{table4}
\end{table}

{As an example, Fig.~\ref{Fig.7} illustrates how the hyperparameter $depth$ impacts the model performance metrics $\sigma_t$ and $\sigma_{v}$. It can be found that, in Fig.~\ref{Fig.7}, the rms deviation $\sigma_{t}$ on the training set exhibits a monotonically decreasing trend with the increasing $depth$ (keeping other default hyperparameters untouched), while on the validation set, the rms deviation $\sigma_{v}$ rapidly decreases at first and then slowly increases, resulting in the appearance of a minimum.} These results are in good agreement with those in Ref.~\cite{Gao2021}. In general, the hyperparameter corresponding to the minimum is referred to as the optimal one. {When opening two or more hyperparameters (namely, let them change freely), the rms deviation, e.g., on the validation set, can be obtained in multi-dimensional hyperparameter space. Indeed, during the practical process of the hyperparameter tuning, one usually need to search the ``optimal'' set of hyperparameters. In Fig.~\ref{Fig.8}, the rms-deviation scatter plots show the training results of the three-dimensional hyperparameter case in the $\sigma_t$ vs $\sigma_v$ plane. One can see that the rms deviations on the testing set for each Skyrme parameter set has been futher improved, decreasing to around 0.2 MeV [even less than 0.2 MeV, e.g., see Fig.~\ref{Fig.8}(d) and (f)] and reaching the accurancy level in the literature~\cite{Gao2021}.} In the case of more than three dimensions, the similar scatter plots can still work well. Based on the selected hyperparameter space and the grid seach method, the optimal hyperparameter sets for different mass tables are presented in Table~\ref{table4}. To avoid the overfitting to some extent, we will use the early stop method, setting $early\_stopping = 30$, to determine the iteration numbers, which is an effective way to detecting the overfitting in the CatBoost algorithm. During the model training-process, the rms deviation on the validation set is always monitored. Once the rms deviation increases for a specified number of iteratinos, the training operation will stop and the $iterations$ at the minimum of the rms deviation will be returned. From the $iterations$ ($< 5000$) in Table~\ref{table4}, it can be noticed that the early stop method does work and stop training in time, which effectively reduces the computing resources and time consumption.

\begin{figure}[htbp]
	\centering
	\includegraphics[width=0.238\textwidth]{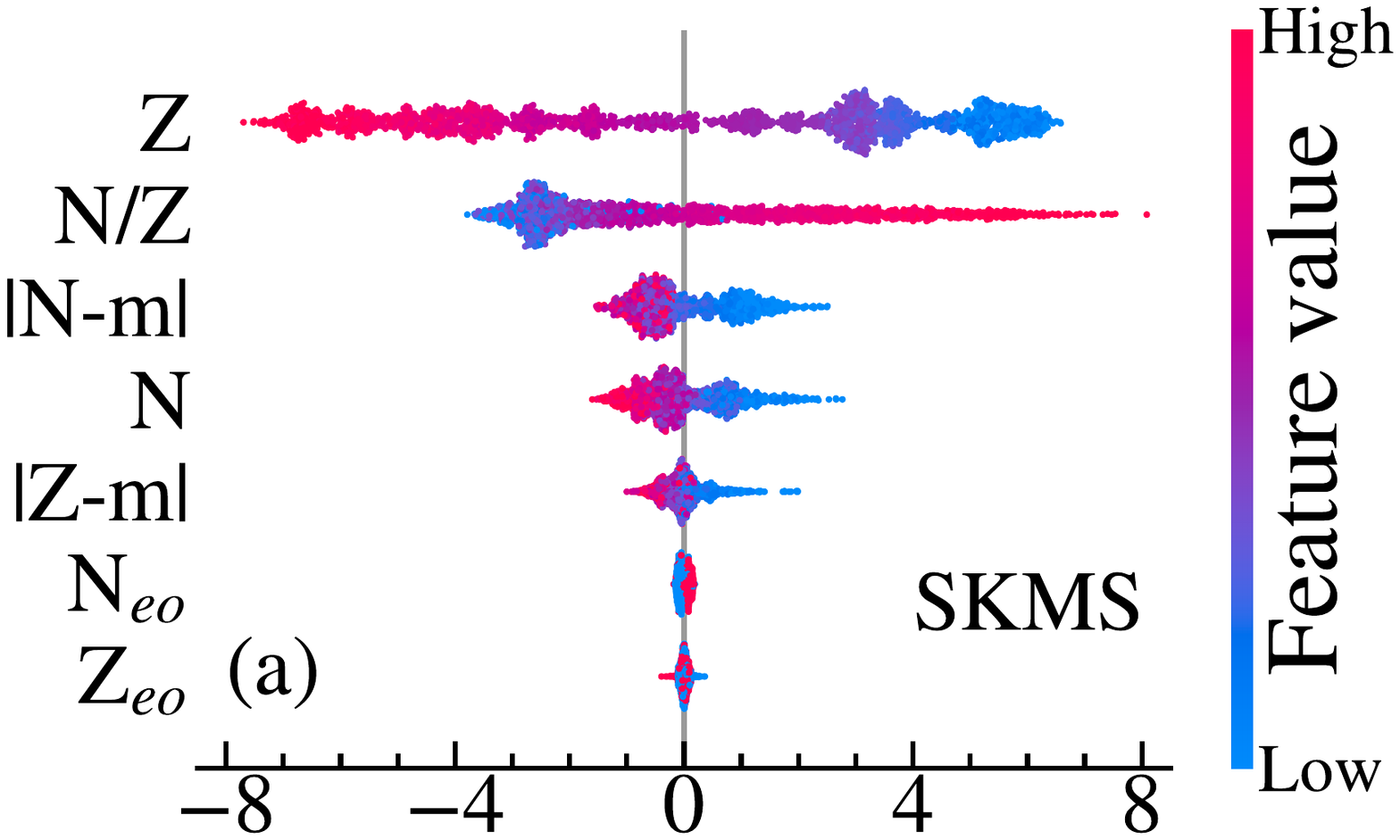}
	\includegraphics[width=0.238\textwidth]{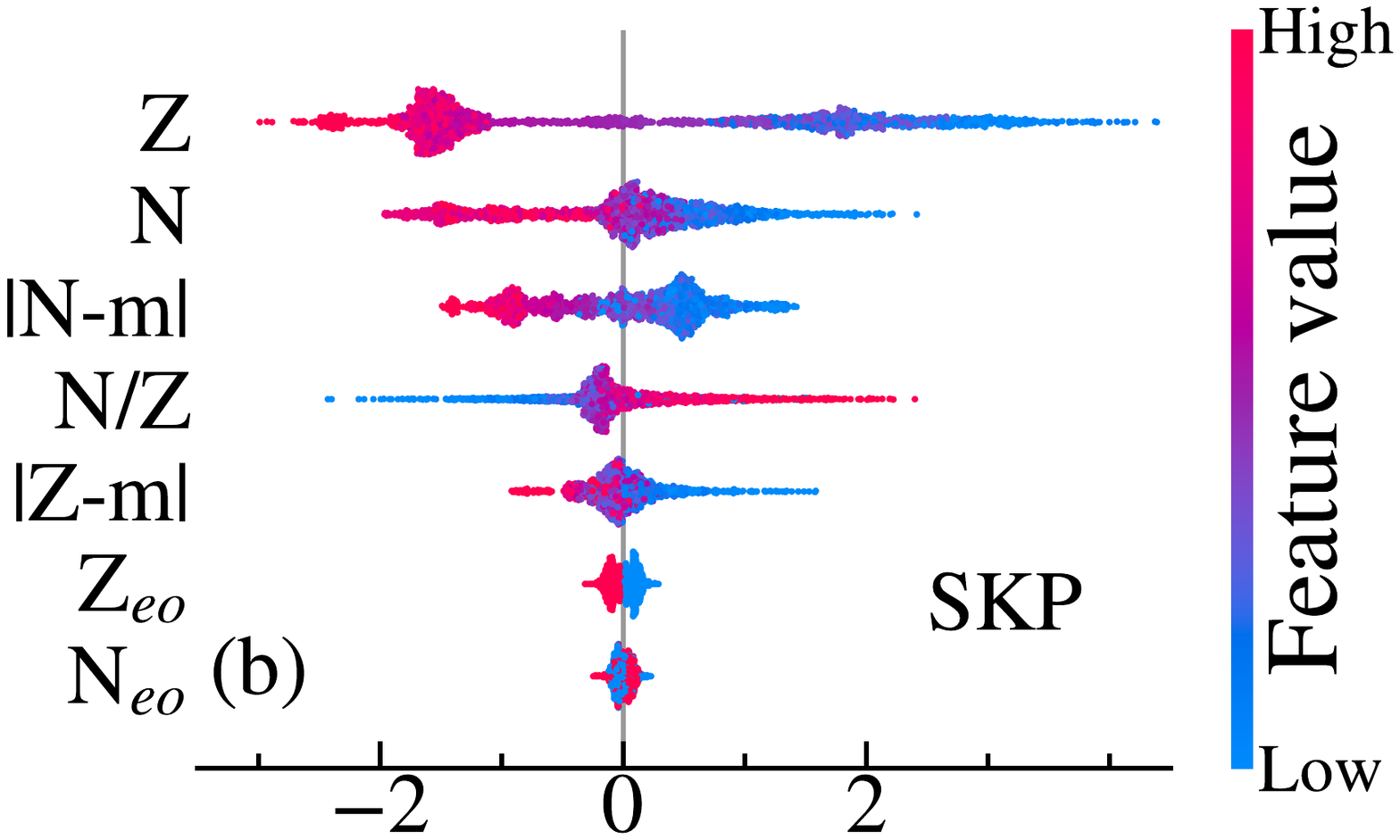}\\
	\includegraphics[width=0.238\textwidth]{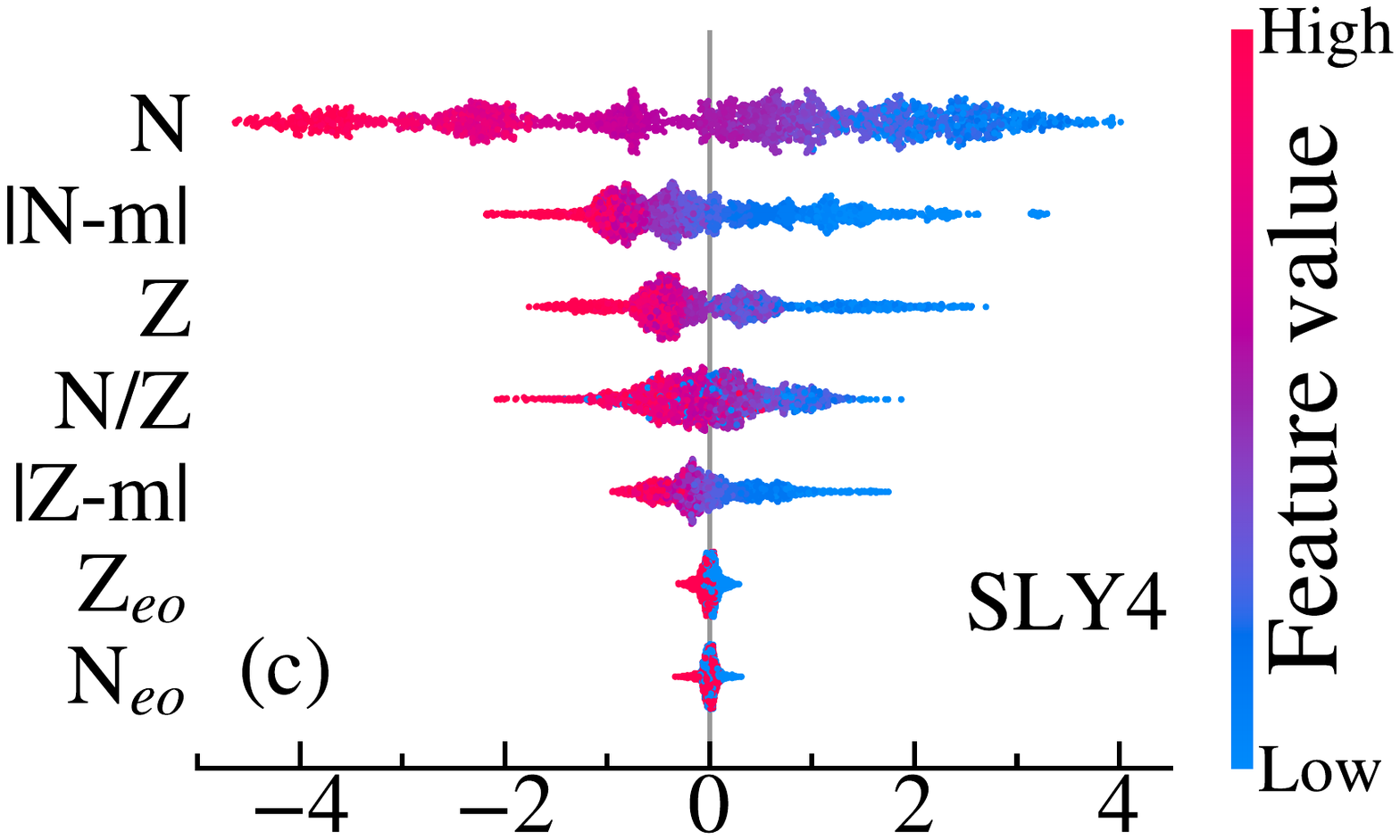}
	\includegraphics[width=0.238\textwidth]{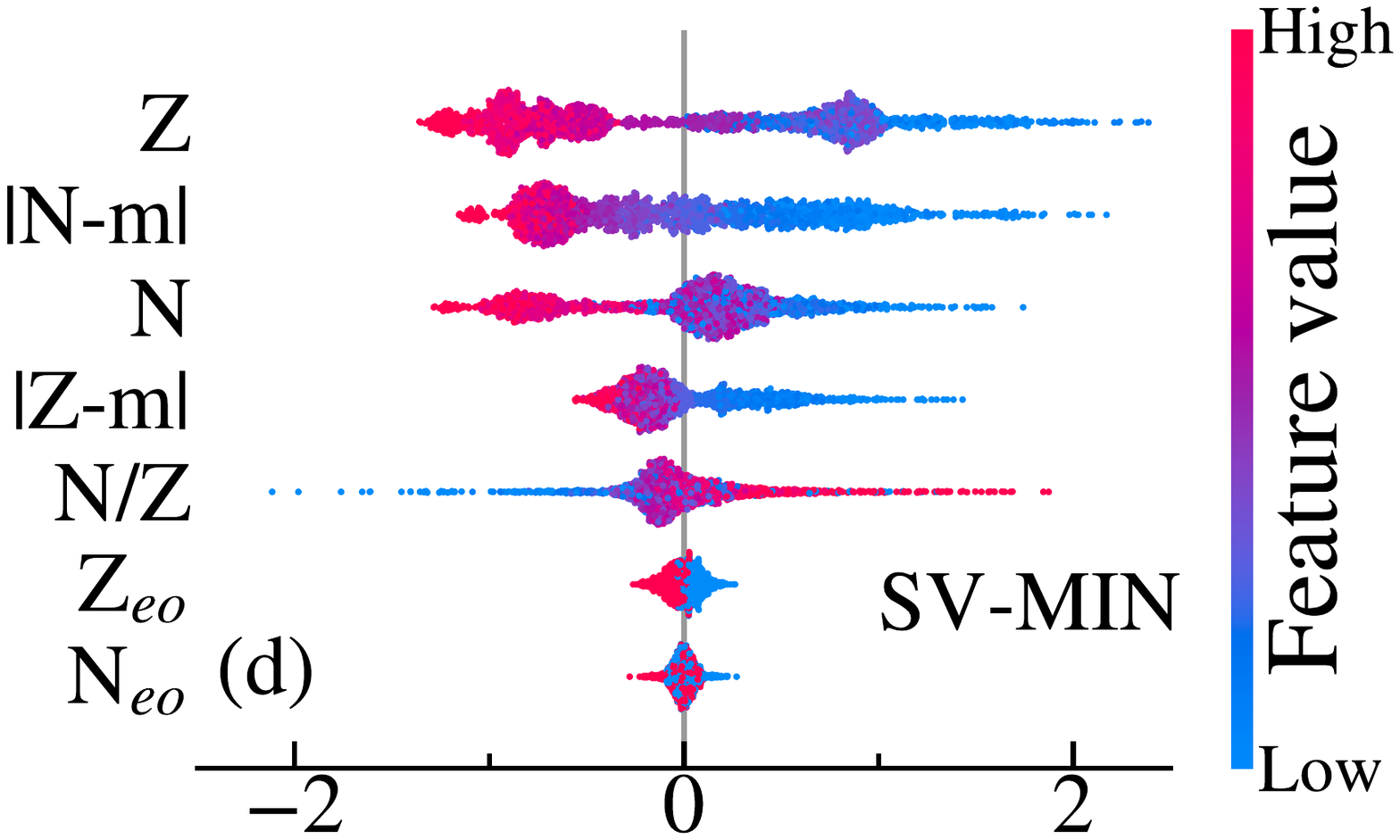}\\
	\includegraphics[width=0.238\textwidth]{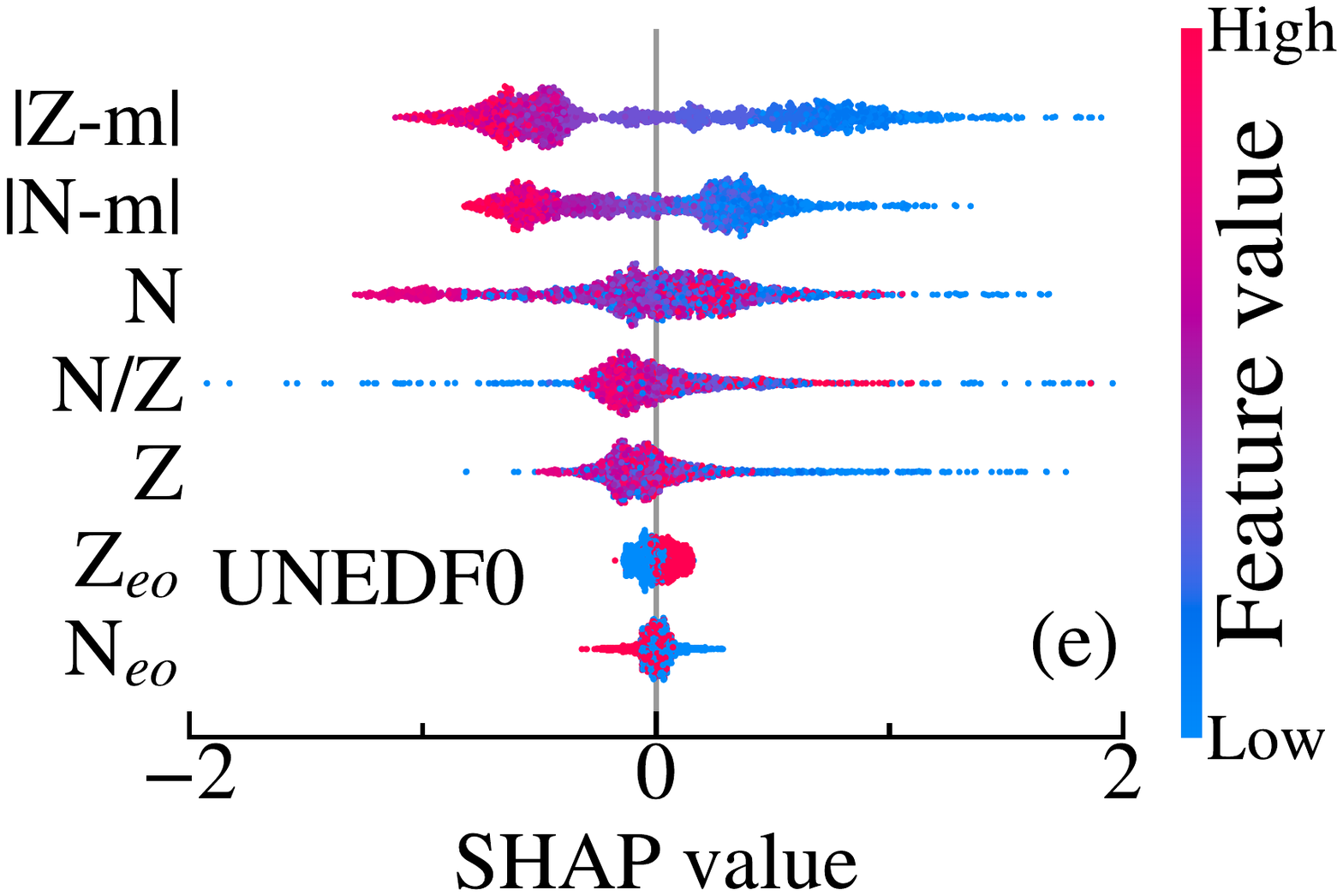}
	\includegraphics[width=0.238\textwidth]{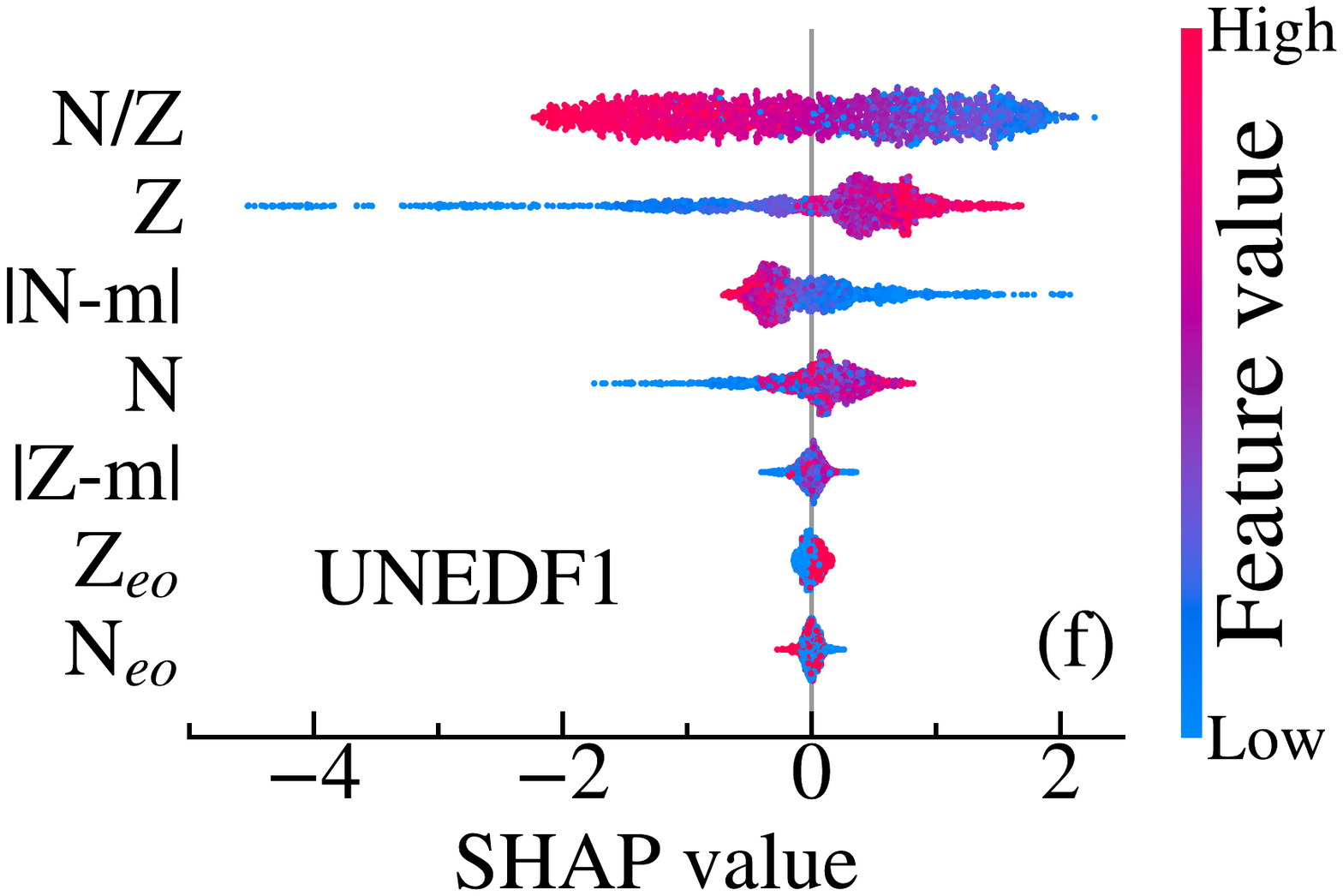}
	\caption{(Color online) Importance ranking for the input features obtained 
		with the SHAP package. Each row represents a feature, and the x-axis
		is the SHAP value, which shows the importance of a feature for a 
		particular prediction. Each point represents a nucleus, and the color 
		represents the feature value (with red being high and blue being low)}
	                                                              \label{Fig.9} 
\end{figure}

\begin{figure}[htbp]
\centering
\includegraphics[width=0.235\textwidth]{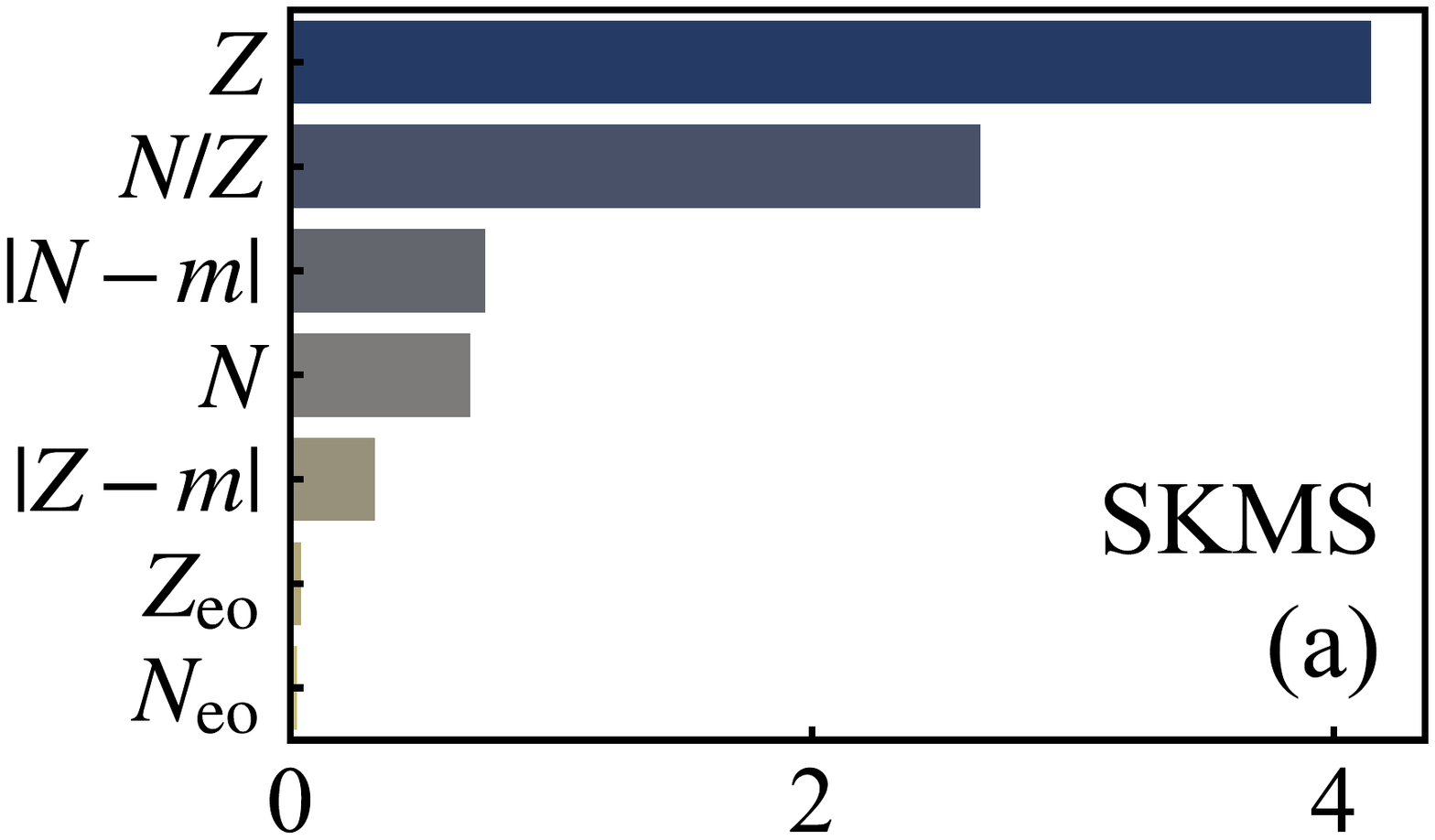}
\vspace{-1.0em}
\includegraphics[width=0.235\textwidth]{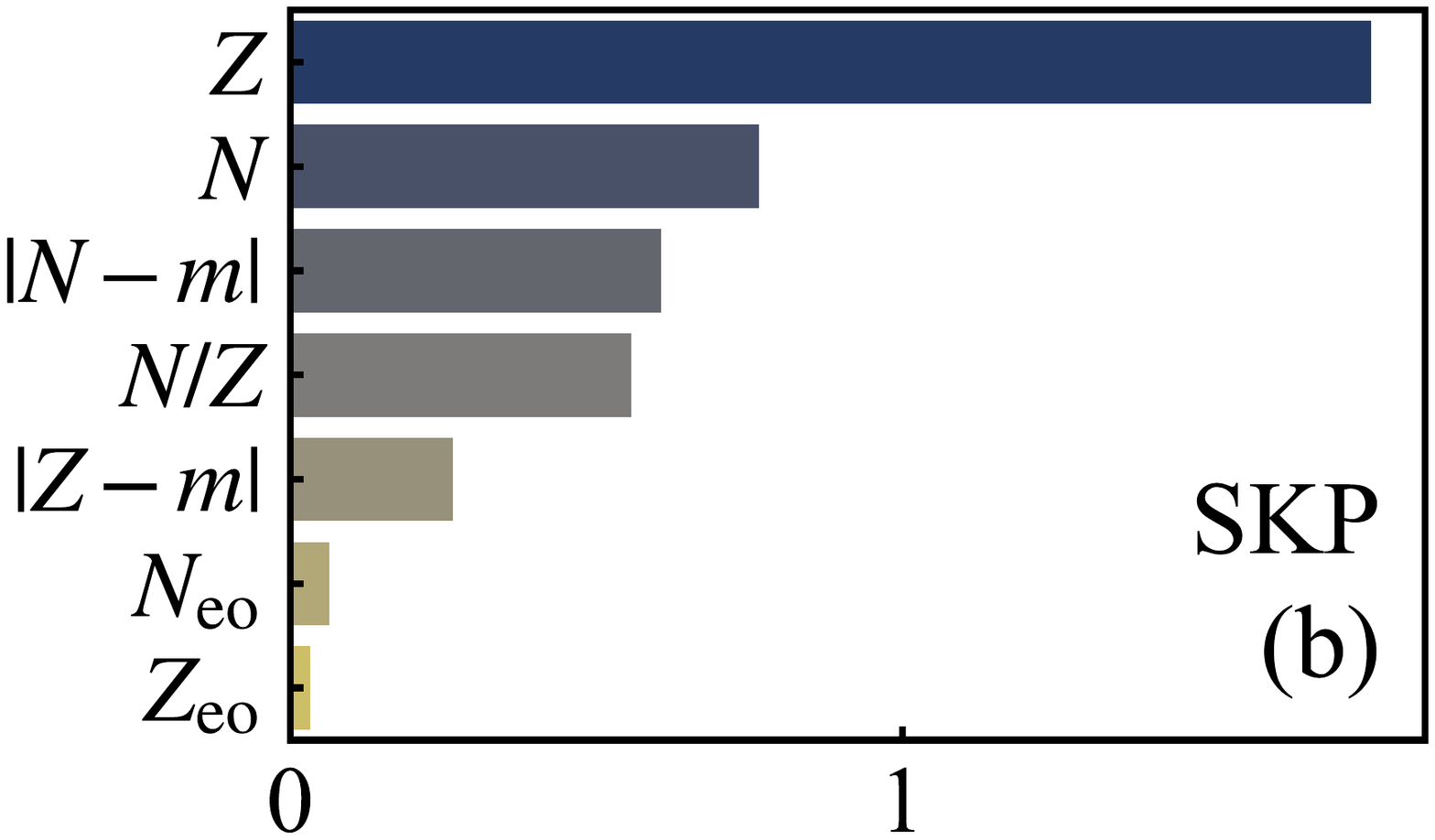}\\
\includegraphics[width=0.235\textwidth]{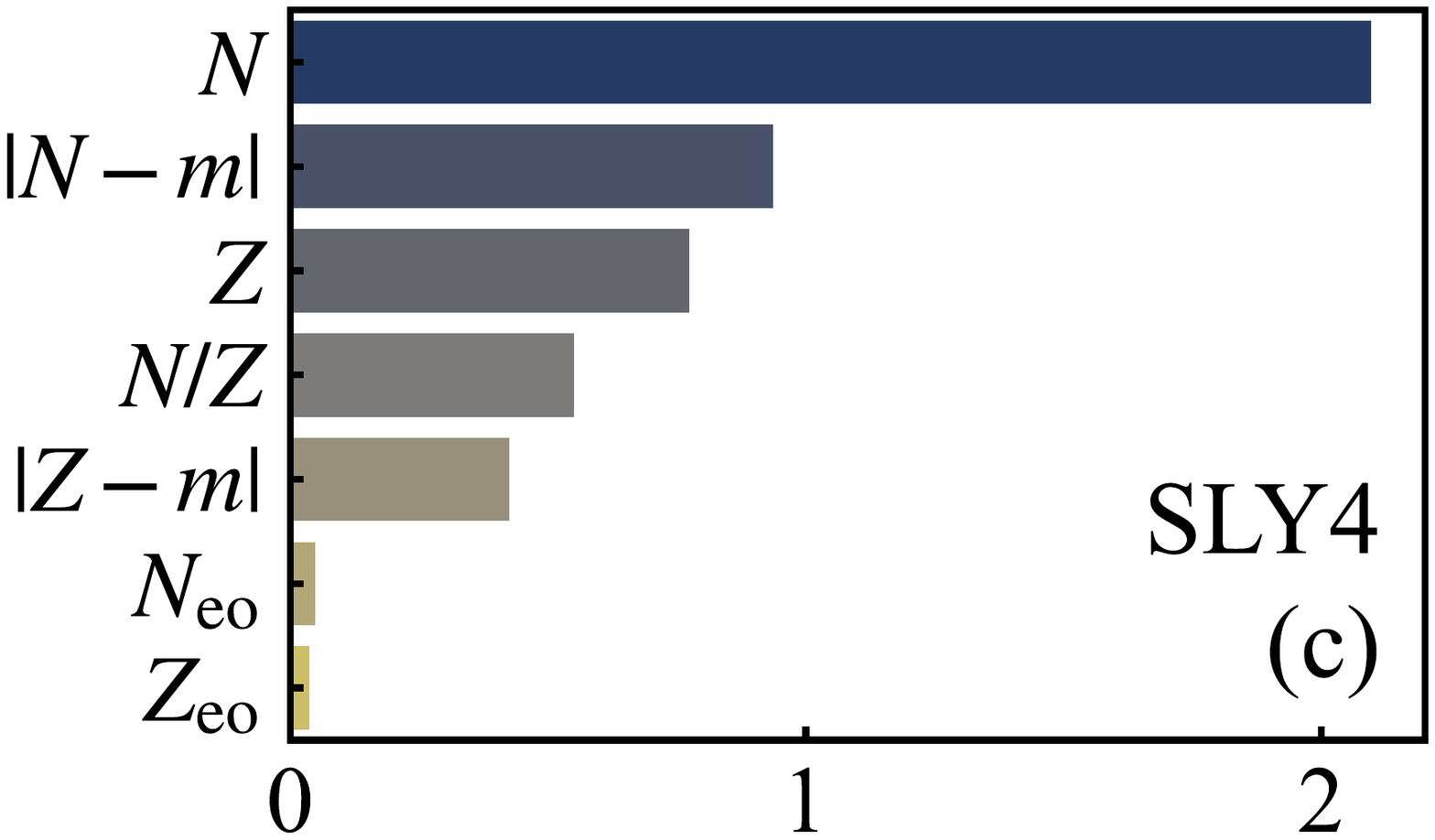}
\vspace{-1.0em}
\includegraphics[width=0.235\textwidth]{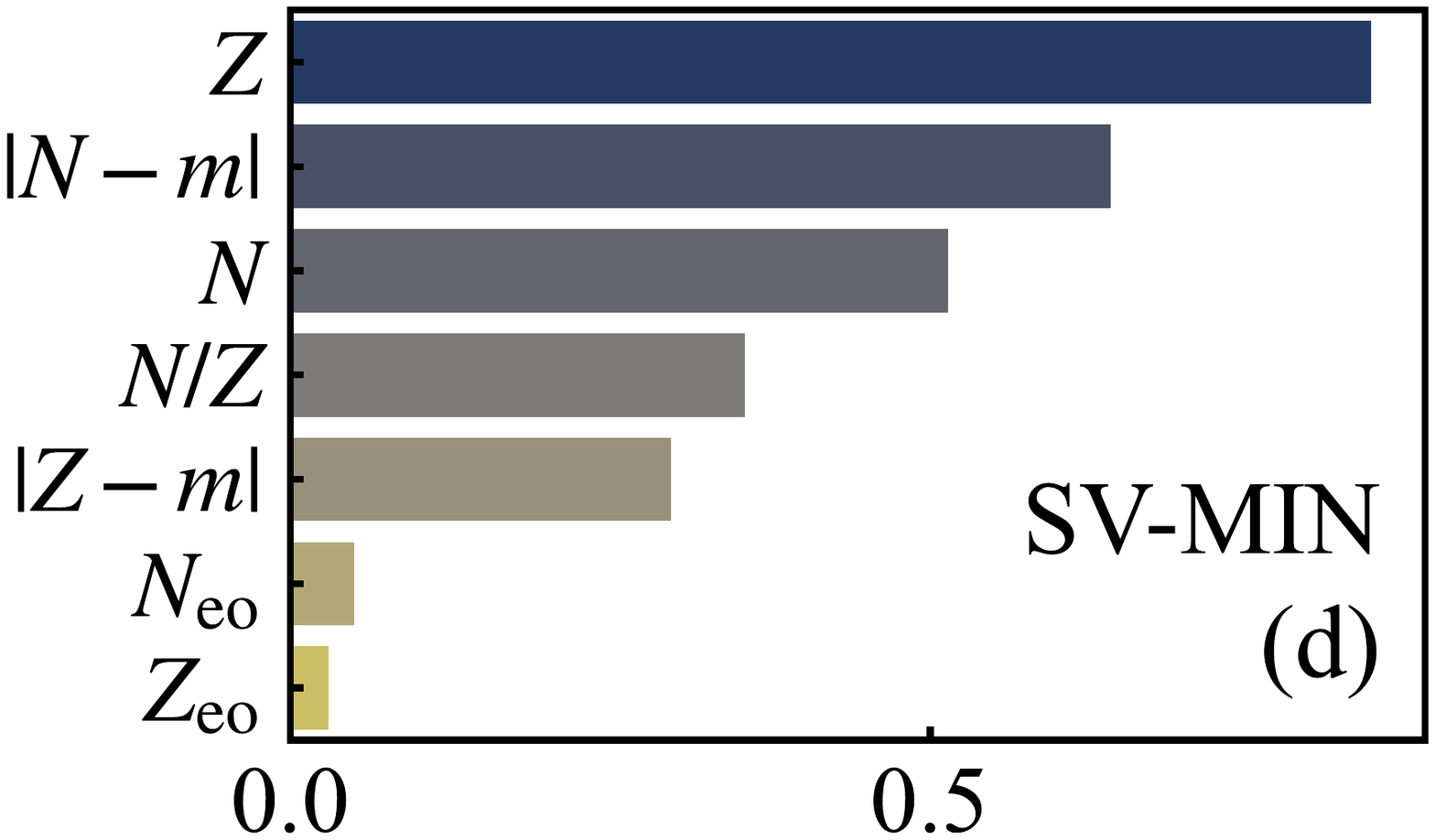}\\
\includegraphics[width=0.235\textwidth]{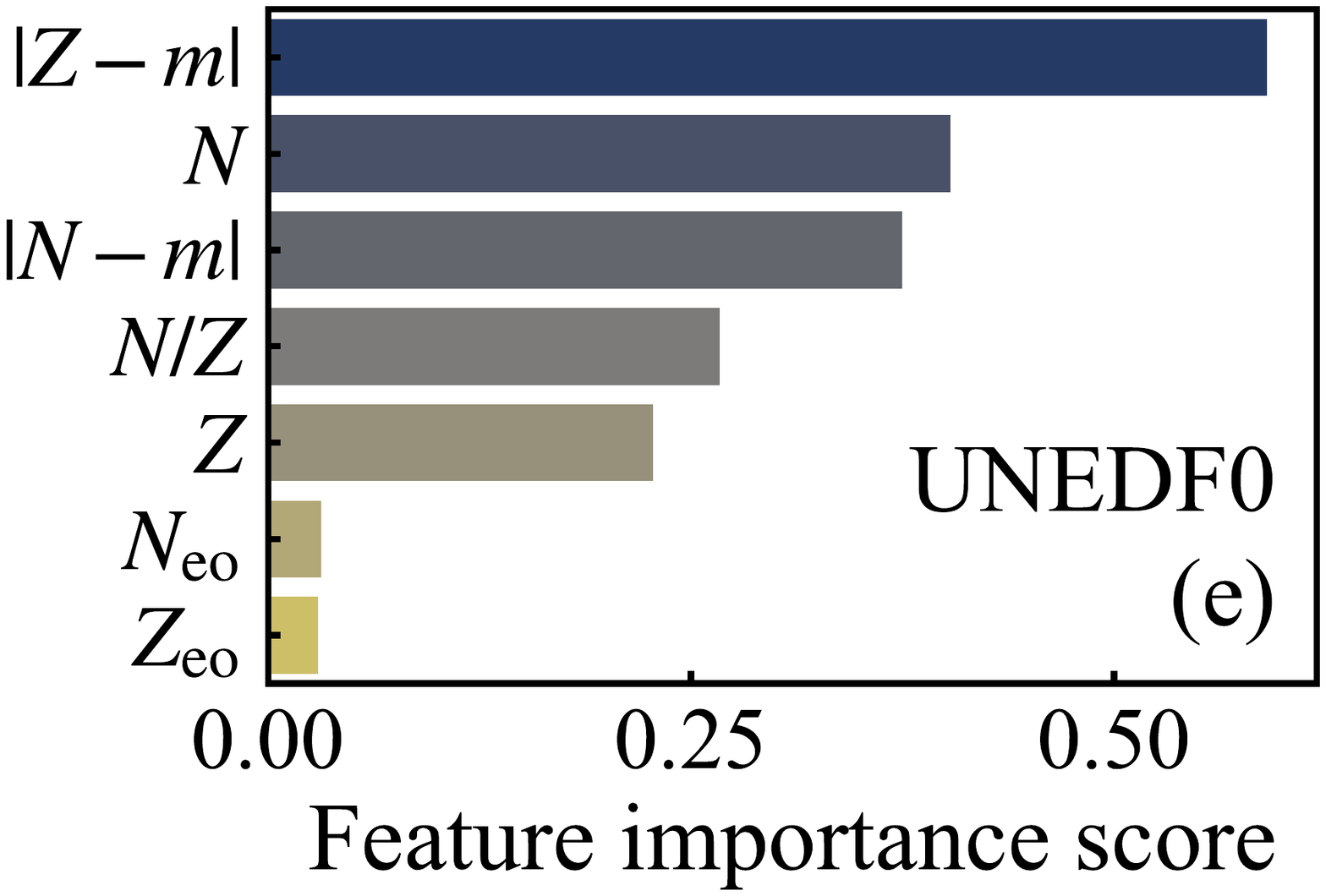}
\includegraphics[width=0.235\textwidth]{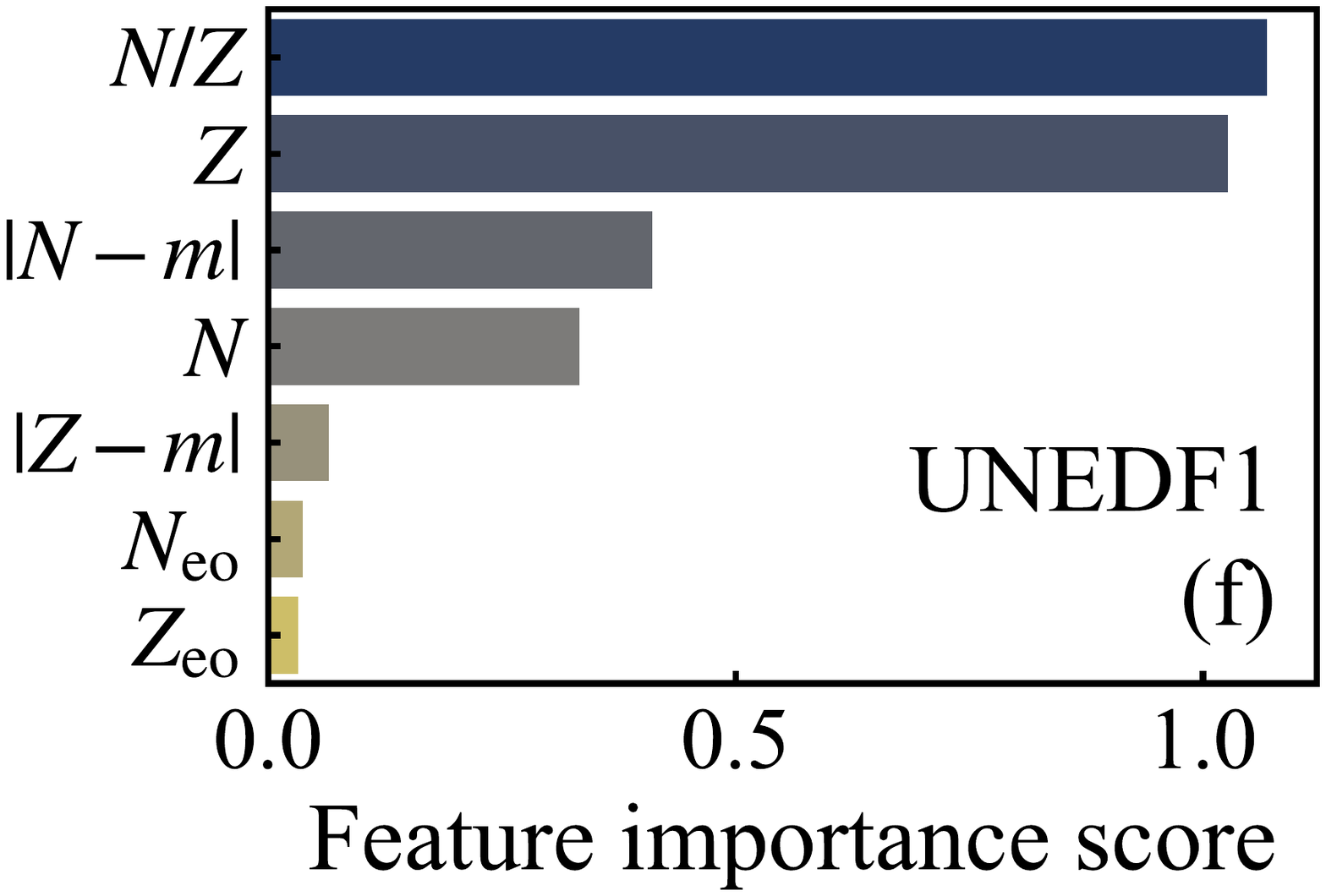}
\caption{(Color online) Feature importance obtained according to the change of the loss function, which represents the difference between the loss value of the model with and without the feature. See text for more details. 
}
														      	  \label{Fig.10}
\end{figure}	


Appreciating the feature importance is instructive to strengthen our understanding on the relationship between the input features and the CatBoost predictions.
As one of the most popular methods, the SHAP (SHapley additive exPlanations)~\cite{lundberg2017unified,ANTWARG2021115736} provides a comprehensive framework for interpreting the contribution of each input feature in the predictions of a machine-learning model, whose key idea is rooted in cooperative game theory and the concept of Shapley values. This approach can quantify the marginal effects of individual features and the interaction/joint effects of two features\cite{ZHOU2016208}. The SHAP should satisfy the properties of local accuracy, missingness, and consistency, e.g., cf. Refs.\cite{VEGAGARCIA2020101039,LIU2023126891} for more details. In the present project, using the optimal hyper-parameter set, the CatBoost algorithm is used to train a regression model based on the training set. Then, the SHAP explainer can be created using the obtained CatBoost model and the SHAP values can be calculated for the testing set. Fig.~\ref{Fig.9} presents the importance ranking for the selected input features. For each Skyrme parameter set, the SHAP values are illustrated in an indivadual subfigure. Such a beeswarm plot can provide a detailed view of feature contributions for every model's prediction. Each point represents a prediction and the plot exhibits the distribution of feature contributions. The top (bottom) feature means the most important (irrelevant) one for predicting the residual between the experimental and theoretical binding energies. Except for the SHAP value, there are some traditional methods to evaluate the feature importance according to how much on average the selected performance metrics (e.g., the prediction value or loss function) changes if the feature value changes~\cite{FeatureImportance}. For a better understanding, we also calculate the feature importance according to the change of the loss function (e.g., the residual rms value), as shown in Fig.~\ref{Fig.10}. It can be found the feature importance is in good agreement with each other between Figs.\ref{Fig.9} and \ref{Fig.10}. For the SkM*, SkP and SV-min parameter sets, the input feature $Z$ is most important. However, the features $N$, $|N-m|$ and $N/Z$ are respectively most important for the SLy4, UNEDF0 and UNEDF1 parameter sets. Also, one can see that the importance ranking for the input features may be different for the CatBoost-refined HFB models with different Skyrme parameters.

\begin{table}[htbp]
	\caption{Based on the corresponding Skyrme parameter set (Par. set), the calculated rms deviations $\sigma_{train}$ and $\sigma_{test}$ (for the training set and testing set, respectively; their standard errors are displayed in parentheses) are listed, along with that of the bare HFB mass model. Except for the last column $R_{MR}$ (model-repair coefficients), all other values are in units of MeV.}
	\renewcommand
	\arraystretch {1.5}
	\centering
	\begin{tabular}{lccccc}
		\hline\hline
	Par. set    &   $\sigma_{train}$  & & $\sigma_{test}$ & $\sigma_{HFB}$ & $R_{MR}$ (\%)\\		
		\hline
        SkM*   & 0.077 (0.002) && 0.259 (0.022) & 7.03 & 96.3 \\	
        SkP    & 0.058 (0.002) && 0.215 (0.021) & 3.55 & 93.9  \\	
        SLy4   & 0.051 (0.001) && 0.215 (0.018) & 5.15 & 95.8 \\	
        SV-min & 0.046 (0.001) && 0.198 (0.019) & 3.33 & 94.1 \\	
        UNEDF0 & 0.083 (0.002) && 0.221 (0.021) & 1.43 & 84.5 \\	
        UNEDF1 & 0.048 (0.001) && 0.189 (0.014) & 1.99 & 90.5 \\	             
		\hline\hline
	\end{tabular}                                          \label{table5}
\end{table}

With the optimal hyperparameter sets as shown in Table~\ref{table4}, the calculated rms deviations and their uncertainties (one standard deviation) are presented in Table~\ref{table5} for the training and testing sets, together with the rms value between expermental data and bare HFB calculations. Supposing that the residual is related to the model deficiency, for decribing the model-repair ability, it is convenient to define a model-repair coefficient by the ratio $R_{MR}=(\sigma_{HFB}-\sigma_{test})/\sigma_{HFB}$ (the subscript $R_{MR}$ indcates the model-repair coefficient, as illustrated in the caption of Table~\ref{table5}). Obviously, such a coefficient can to some extent represent the ability that the machine-learning algorithm captures the law in the residual data. In general, it can be said that the machine-learning algorithm possesses model-repair ability when $0 < R_{MR} \leq 1$ while has not when $R_{MR} \leq 0$. Ideally, the model is entirely repaired by the machine-learning model when $R_{MR}$ equals one, which means that the missing physics may already be fully recovered through machine learning. For a given physically-based model, needless to say, the larger the coefficient $R_{MR}$ is, the stronger the model-repair ability of the machine-learning algorithm is. The model-repair coefficients are calculated, as given in Table~\ref{table5}, for six different Skyrme parameter sets used here. Indeed, significant improvements of approximately 96.3\%, 93.9\%, 95.8\%, 94.1\%, 84.5\% and 90.5\% after the CatBoost refinement on the SkM*, SkP, SLy4, SV-min UNEDF0 and UNEDF1 
are obtained, indicating the strong capability of CatBoost to improve 
theoretical nuclear mass models. We can see that the CatBoost method exhibits different model-repair abilities for different Skyrme-parameter sets. It should be pointed out that in different physically-based models, the comparison of the model-repair coefficients is somewhat meaningless since these models, in principle, miss different ``physics''. 

\begin{figure}[htbp]
\includegraphics[width=1.0\linewidth]{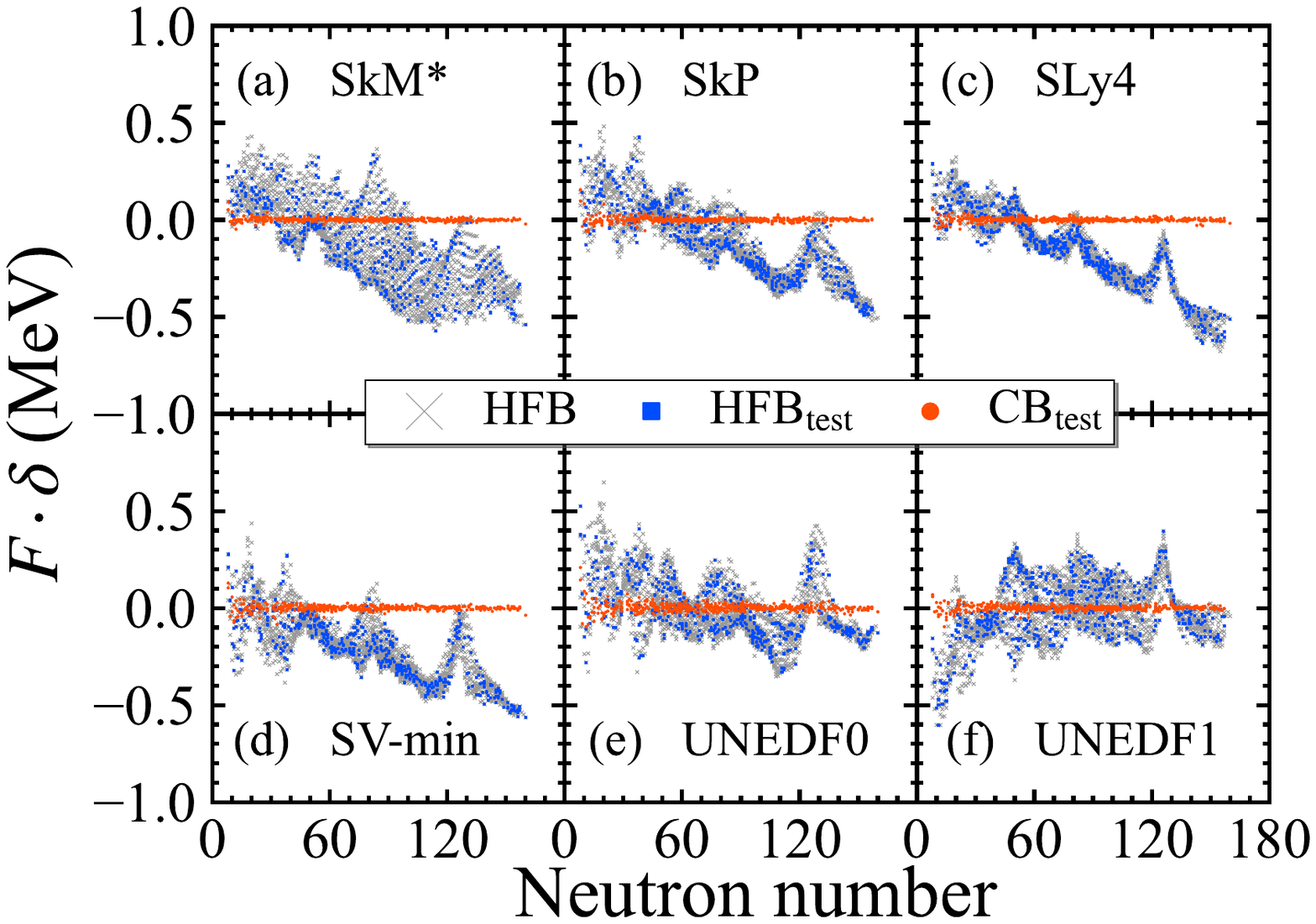}
\caption{(Color online) Residual $\delta$ = B$_{\mathrm{th}}$(Z,A)-B$_{\mathrm{exp}}$(Z,A) plotted as a function of neutron number. Gray-cross symbols represent the residuals obtained from the bare HFB dataset with the corresponding Skyrme parameters; Blue diamonds indicate the selected data for the testing set and the solid red-circles illustrate the residuals obtained with the CatBoost-refined HFB models. For comparison, the residual data are uniformly scaled by the factor $F=1/(\delta_{max}-\delta_{min})$ for all the datasets with different Skyrme parameters. See text for more details.}
																  \label{Fig.11}
\end{figure}

As it is well known, any physics model is not the physics truth itself and its uncertainty of prediction can be reduced but impossible to be eliminated. If the mathematical structure of the model is suitable, its predicted uncertainty should primarily originate from the uncertainty propagation of the model parameters. For a good physics-model, it is expected to possess the high predicting capability and small model bias. Fig.~\ref{Fig.11} illustrates the residuals between experimental data and theoretical results calculated by the bare and refined mass models. Such scatter plots can help displaying the distribution properties of the residuals as well as the model predictive abilities. For a more ideal model, the residuals should be homoscedastic (i.e. the variance of the residuals should be constant in different regions) along some degree of freedom such as the nucleon number and the binding energy. In Fig.~\ref{Fig.11}, we arbitrarily plot the scaled residual $F\cdot\delta$ in a function the neutron number (similarly, for other variables), where $F$ [$\equiv$ 1/($\delta_{max}-\delta_{min}$)] is the scaling factor. It can be seen that, prior to machine learning, the residual distribution presents the nonrandom distributed shape. After refining by the CatBoost algorithm, not only the rms deviations significently decrease, but also the residual distributions become more random and the specific patterns disappear to a large extent, indicating that the machine-learning model pick up the missing physics in the bare mass models to some extent.

\begin{figure}[htbp]
\centering
\includegraphics[width=1.0\linewidth]{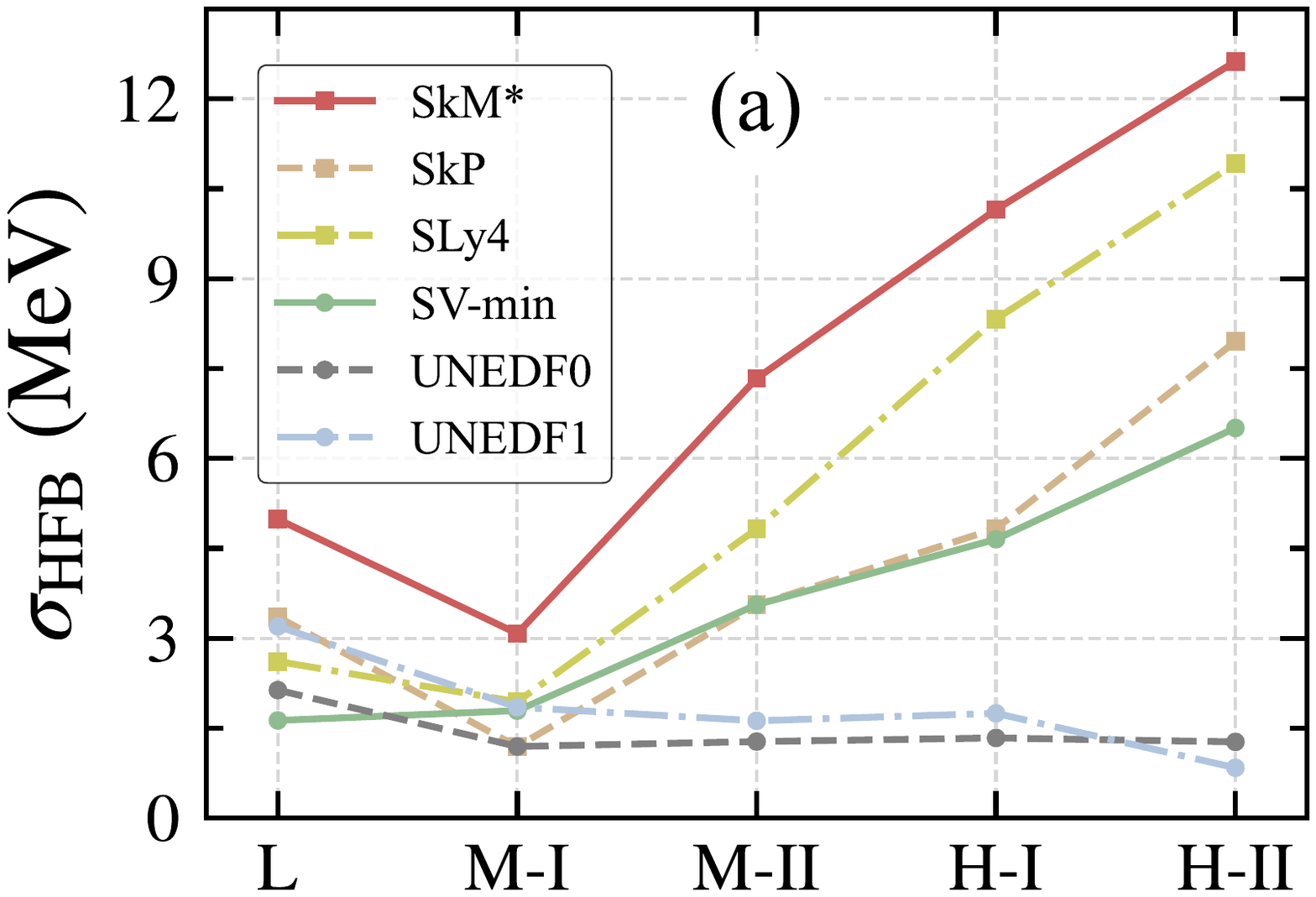}
\includegraphics[width=1.0\linewidth]{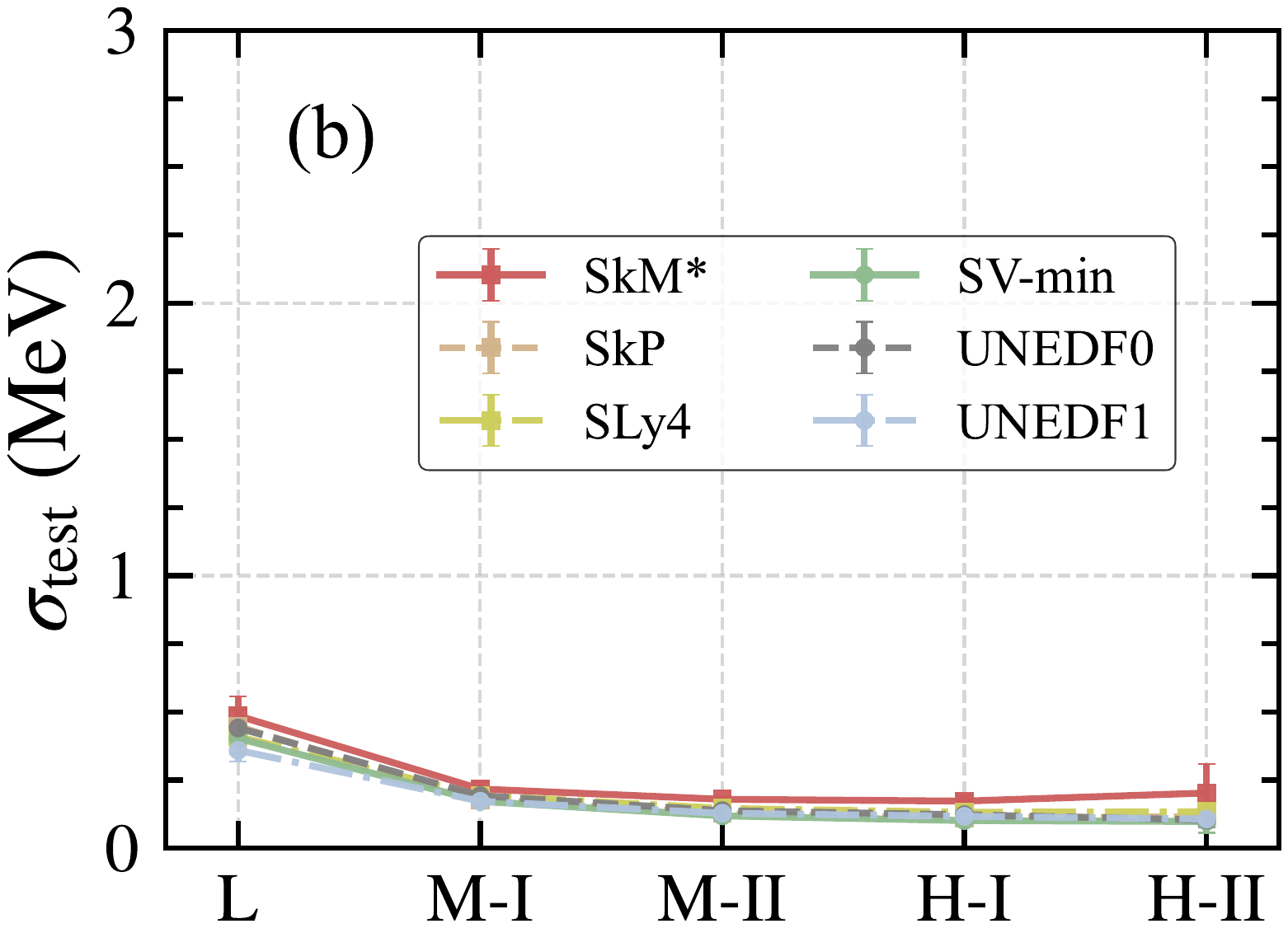}
\caption{(Color online) (a) The rms deviations $\sigma_{HFB}$ obtained with the bare HFB models with six Skyrme parameter sets for light (L: $8\le Z<28$), medium-I (M-I: $28\le Z<50$),	medium-II (M-II: $50\le Z<82$ ), heavy-I (H-I: $82\le Z<100$) and heavy-II (H-II: $Z\ge100$ ) mass regions. (b) Similar to (a), but for the testing set obtained with the CatBoost-refined HFB mass models. For more details, see the text.
}
	                                                              \label{Fig.12}
\end{figure}
\begin{figure}[htbp]
\includegraphics[width=1.0\linewidth]{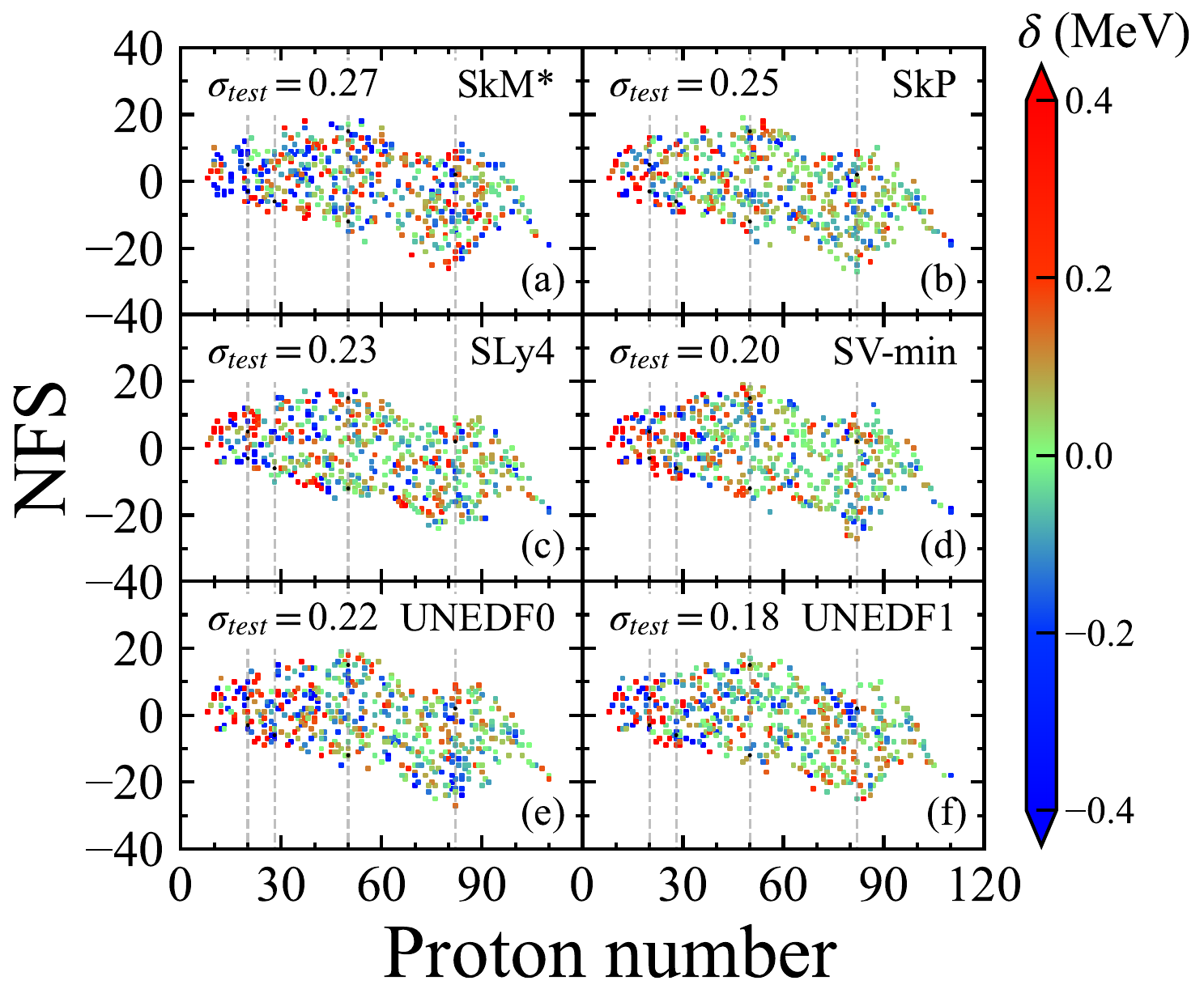}
\caption{(Color online) Similr to Fig. \ref{Fig.2}, but for the predicted residuals by CatBoost-refined HFB models on the testing set.}
																  \label{Fig.13}
\end{figure}

From Fig.~\ref{Fig.11}, it can be noticed that the variance depends on the nuclear mass regions, indicating the appearance of heteroscedasticity. For instance, it seems that the residuals have the overall decreasing trend, from Fig.~\ref{Fig.11}(a) to (d), with the increasing neutron number. In order to further compare the predicted accuracies in different nuclear domains, e.g., from light to heavy mass regions (similar to those in Refs.~\cite{PhysRevC.89.024311,PhysRevC.90.017302, SOBICZEWSKI20181}),  we divided the the experimental dataset collected in AME2020 with $Z, N \ge 8$ into light $(8\le Z<28)$, medium-I $(28\le Z<50)$, medium-II $(50\le Z<82)$, heavy-I $(82\le Z<100)$ and heavy-II $(Z\ge100)$ regions. In each mass region, the rms deviations obtained from the bare and CatBoost-refined HFB calculations labeled by different Skyrme parameter sets are illustrated in Fig.~\ref{Fig.12}. It can be seen that the rms deviations predicted by the bare HFB models strongly depend on the nuclear mass regions, from light to heavy-II regions, exhibiting an increasing trend except for the cases labeled by UNEDF0 and UNEDF1. Even, the differences of the rms deviations between the smallest (e.g., at the M-I region) and largest values reach around 10 MeV for the bare HFB predictions with e.g., SkM* and SLy4 parameters. For the CatBoost-refined HFB predictions, as seen in Fig.~\ref{Fig.12}(b), both the rms deviations and their dependence on nuclear mass regions decrease significantly, indicating that the deficiencies of the bare HFB models are repaired by the CatBoost algorithm to a large extent. Note that the data in Fig.~\ref{Fig.12}(b) are obtained by 500 repetitions, similar to the operations of Fig.~\ref{Fig.4} but using the optimal hyper-parameter sets. {Further, Figs.~\ref{Fig.13} illustrates the detailed distributions of the mass residuals for the testing set in $Z-NFS$ planes, which can clearly exhibit their uniform properties.} The results are obtained by randomly spliting the entire dataset into the training and testing sets at the ratio 4:1 once. During the training process, the so-called optimal hyper-parameters as illustrated in Table~\ref{table5} are adopted. From such perspectives, it can be found that there is no obvious distributed-patterns. Namely, the residuals on the selected testing set display a relatively random distribution. Moreover, the one-time rms deviations are already reduced to 0.27, 0.25, 0.23, 0.20, 0.22 and 0.18 MeV, which are close to those average values of 500 runs (see, e.g., Table~\ref{table5}), indicating the stability of the predictions. These facts mean that the machine-learning algorithm CatBoost can capture the missing physics of the corresponding Skyrme force and decode the correlation between the input features and the residuals.

\begin{figure}[htbp]
\centering
\includegraphics[width=1.0\linewidth]{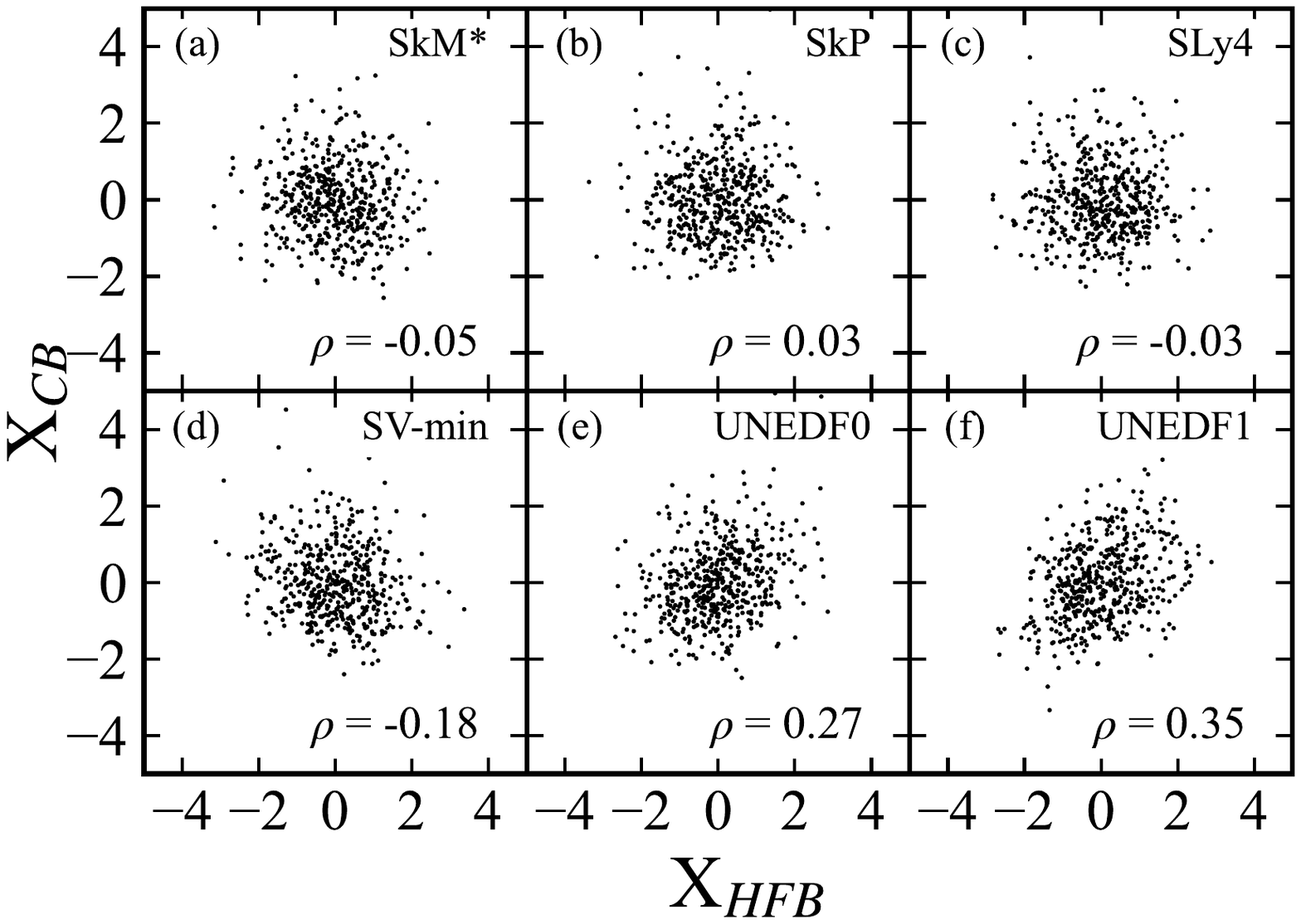}
\caption{(Color online) Two-dimensional scatter plots of the rms deviations between the results obtained by the bare and CatBoost-refined HFB model on the testing sets. Data is scaled by using the $z-$score normalization.}
															      \label{Fig.14}
\end{figure}

Indeed, one can imagine that, if the CatBoost algorithm can capture the physics law hidden in the residual data in a real sense, the prediction of the refined HFB mass model will not depend on the results obtained by the bare HFB calculations. Naturally, it is somewhat necessary to know whether or not the larger the rms deviation of the bare HFB model is, the larger that of the refined-HFB model will be. As it is well known, a scatter plots between two quantities can be helpful to visualize and determine this kind of relationships or correlations. Fig.~\ref{Fig.14} illustrates such plots of the rms deviations between the bare and CatBoost-refined HFB models, together with the correlation coefficients $\rho$. For display purpose only, using the data scaling technique, e.g., the standardization method, we transform the rms deviations for each repetition (500 times in total) by the $z$-score normalization formula $X_i=(\sigma_i-\bar{\sigma_i})/\sigma_s$, where $\sigma_i$, $\bar {\sigma_i}$ and $\sigma_s$ mean the corresponding rms deviation, the mean value of the 500 rms deviations and the standard deviation of these rms deviations, where the index $i$ denotes the CatBoost-refined (CB) or bare HFB models. At this moment, the quantity $X_i$ is dimensionless and if two quantities, e.g., $X_{CB}$ and $X_{HFB}$, are independent each other, the shape of the scatter plot will be approximately round. In Fig.~\ref{Fig.14}, it can be noticed that the rms distributions almostly exhibit a round shape for every Skyrme parameter sets. To measure the direction and the correlation strength of two quantities, the Pearson correlation coefficient $\rho$, cf. e.g., Ref.~\cite{Zhang2021}, is also shown in this figure.  From Fig.~\ref{Fig.14}, we can also see that these two quantities have the positive (negative) correlations for six datasets. However all the correlations are very weak according to a so-called thumb rule which usually considers the correlation is strong (moderate, weak and very weak or none) when the absolute value of the correlation coefficient is larger than 0.7 (between 0.7 and 0.5, between 0.5 and 0.3, and smaller than 0.3). Except for the case in Fig.~\ref{Fig.14}(f) (the correlation coefficient $\rho=0.35$), these plots display the relationships with very weak correlations ($|\rho| < 0.3$). The predictions of the refined mass models hardly depend on the data obtained by the bare physics-models, indicating the performance property of the relatively ``good'' model.

\begin{figure}[htbp]
\centering
\includegraphics[width=1.0\linewidth]{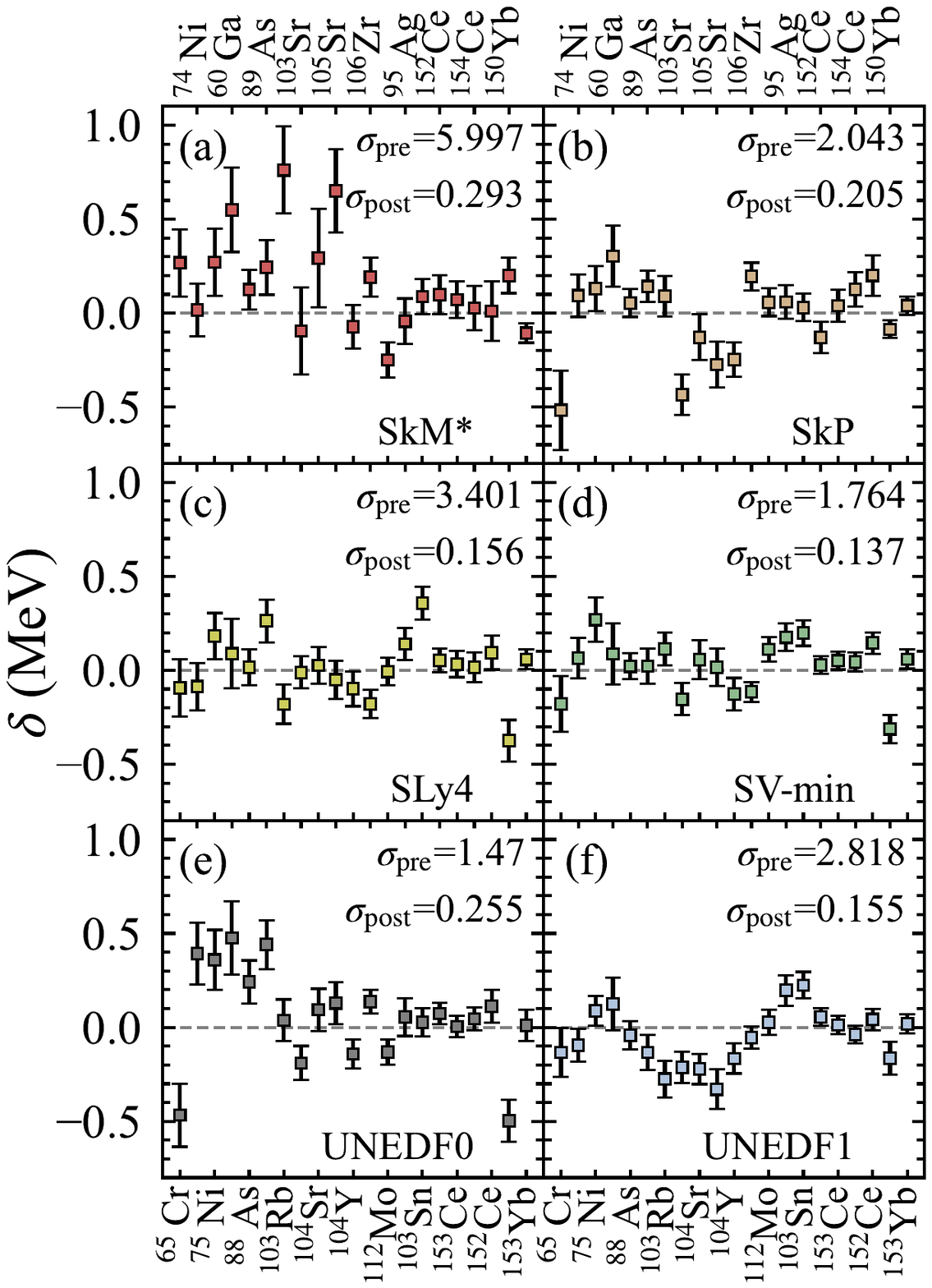}
\caption{(Color online) Residuals $\delta(Z, N) = B_{th} -B_{exp}$ between the theoretical (predicted by the CatBoost-refined HFB calculations) and experimental binding energies for 21 newly measured nuclei (outside AME2020). Data taken from Refs.\cite{SILWAL2022137288,GIRAUD2022137309,PhysRevC.104.065803,PhysRevC.109.035804,PhysRevC.103.044320,PhysRevC.103.044320,hukkanen2024precision,hukkanen2024precision,hukkanen2024precision, ge2024highprecision,PhysRevC.107.014304,PhysRevC.105.L052802,kimura2024comprehensive,PhysRevLett.127.112501}.  See the text for more details.}
															      \label{Fig.15}
\end{figure}

In recent years, precision mass measurements for some nuclei outside AME2020 have been performed, including $^{65}$Cr\cite{SILWAL2022137288}, $^{74-75}$Ni\cite{GIRAUD2022137309}, $^{60}$Ga\cite{PhysRevC.104.065803}, $^{88-89}$As\cite{PhysRevC.109.035804}, $^{103}$Rb\cite{PhysRevC.103.044320}, $^{103-105}$Sr\cite{PhysRevC.103.044320}, $^{104}$Y\cite{hukkanen2024precision}, $^{106}$Zr\cite{hukkanen2024precision}, $^{112}$Mo\cite{hukkanen2024precision}, $^{95}$Ag\cite{ge2024highprecision}, $^{103}$Sn\cite{PhysRevC.107.014304}, $^{152-155}$Ce\cite{PhysRevC.105.L052802,kimura2024comprehensive} and $^{150,153}$Yb\cite{PhysRevLett.127.112501}. To test the prediction ability of the CatBoost-refined HFB models with different Skyrme forces on these newly measured members across the nuclear chart, Fig.~\ref{Fig.15} presents the binding-energy residuals for different HFB mass datasets, together with the rms deviations $\sigma_{pre}$ and $\sigma_{post}$ which are respectively based on the bare and refined HFB calculations. The error bars for these 21 nuclei correspond the standard deviation of the residuals obtained by 500-time repetitions of the model predictions, illustrating the predicted stability. Each time, the CatBoost modeling and model prediction are respectively performed by randomly selecting 80\% of the dataset of AME2020 (with $Z,N >8$) as the training set and these 21 nuclei as a testing set. From Fig.~\ref{Fig.15}, one can find that, comparing the present results with those in Table~\ref{table5}, the CatBoost-refined HFB predictions for the rms deviations and the residuals are highly successful, possessing the satisfactory accurancies.  

\begin{figure}[htbp]
	\centering
	\includegraphics[width=0.95\linewidth]{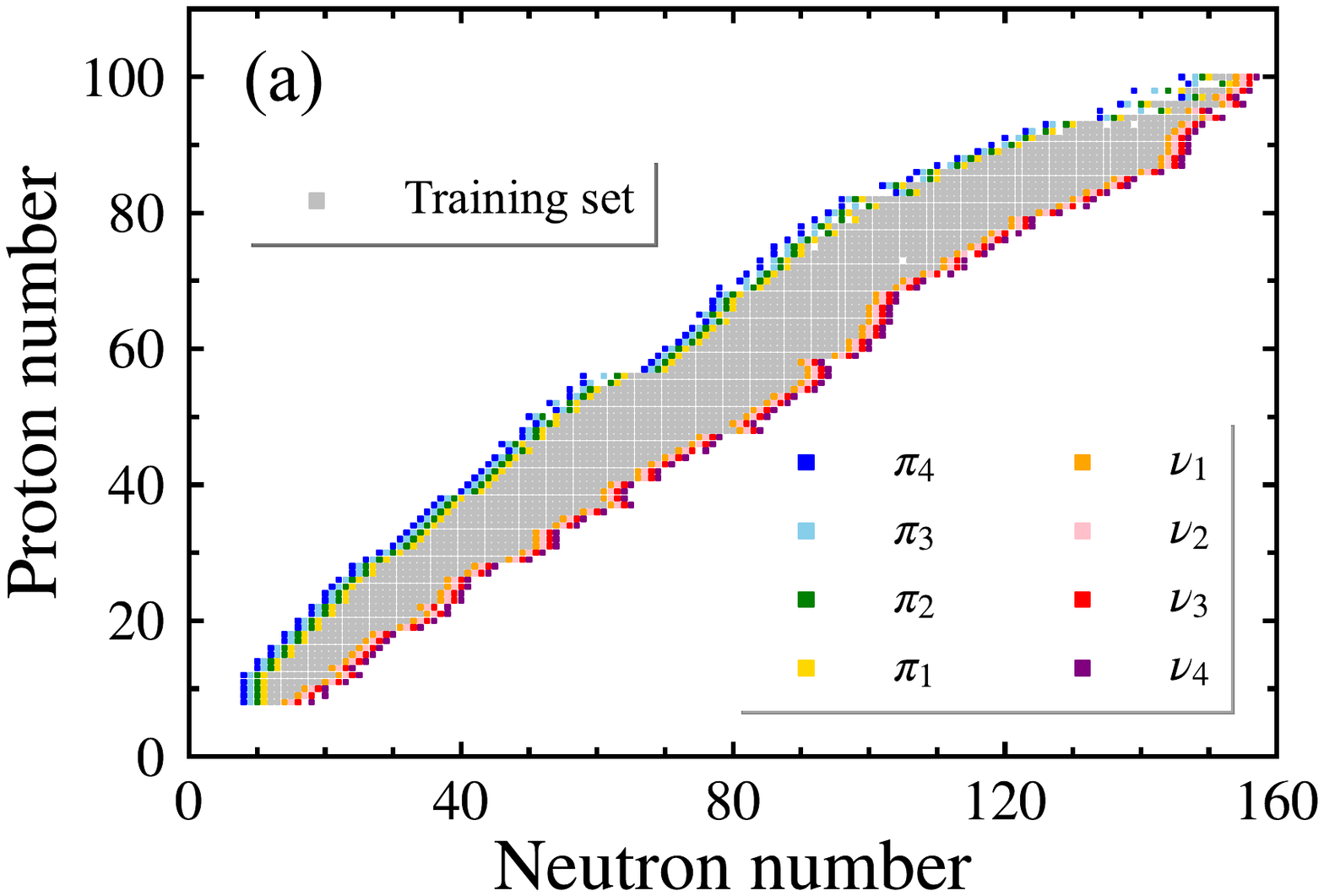}\\
	\includegraphics[width=0.95\linewidth]{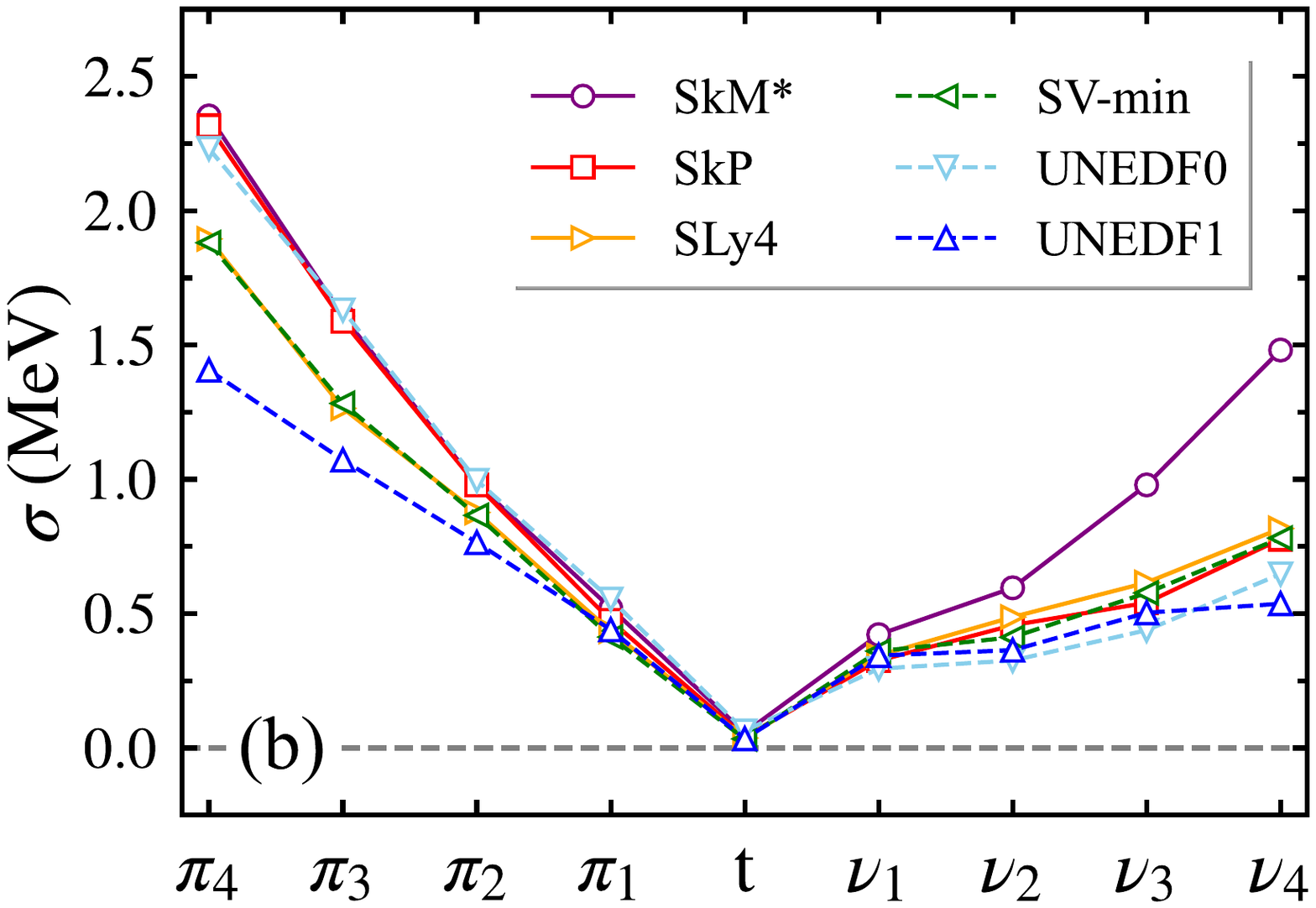}
	\caption{(Color online) (a) The selected nuclei in the training set (gray squares) and eight testing sets (colored squares) for examining the extrapolation power for drip-line nuclei. The testing sets on the proton-rich and neutron-rich sides are respectively denoted by $\pi_{\mathrm{i}}$ and $\nu_{\mathrm{i}}$, $i=1,2,3,4$.  (b) Comparison of the extrapolation power of the CatBoost-refined HFB models with different Skyrme parameter sets for eight testing sets. The symbol $t$ indicates the training set. See text for more details.}
															      \label{Fig.16}
\end{figure}

\begin{figure}[htbp]
	\centering
	\includegraphics[width=1.0\linewidth]{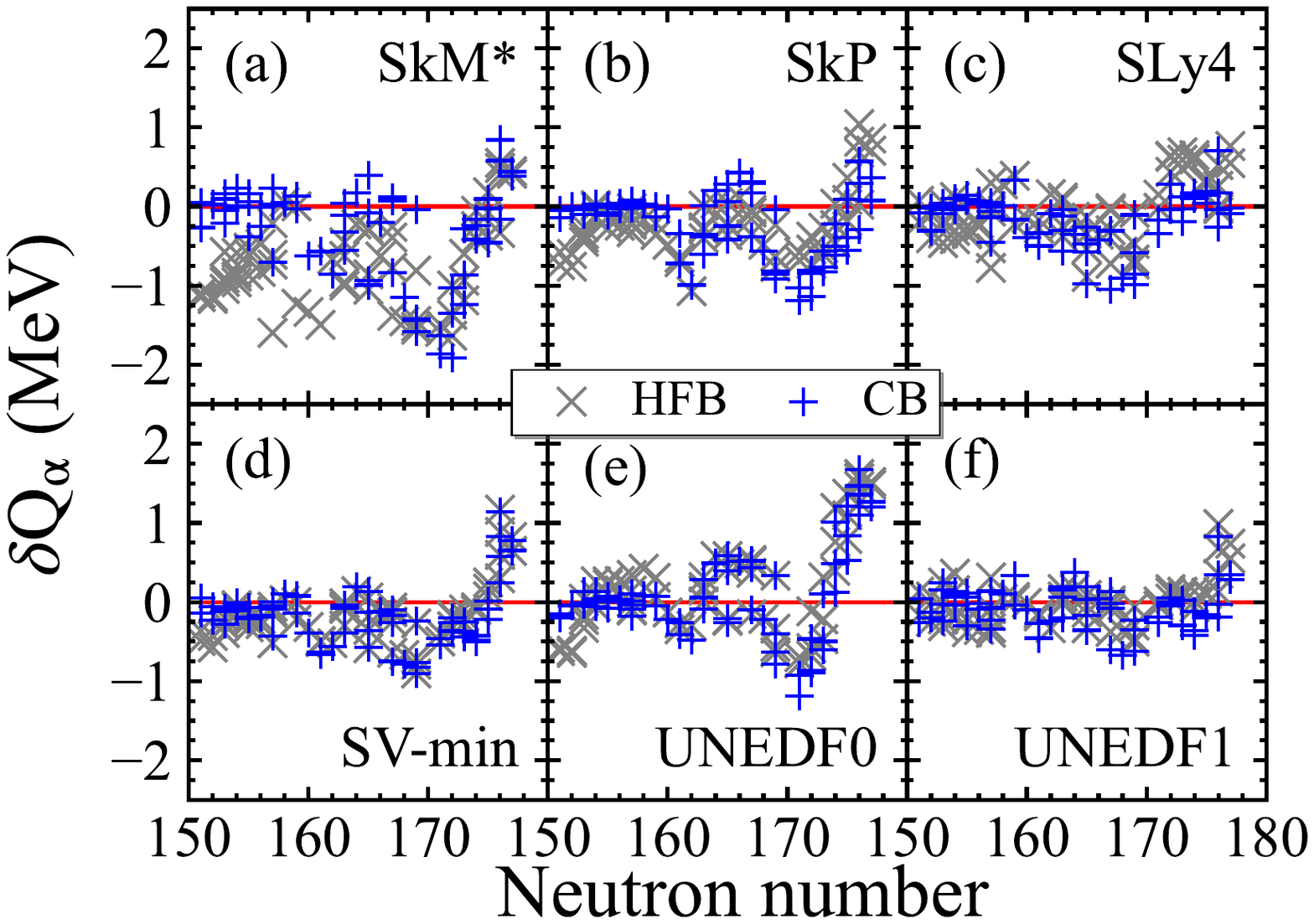}
	\caption{(Color online) Residual $\delta$Q$_{\rm \alpha}$($\equiv Q_{\alpha}^{\rm exp} - Q_{\alpha}^{\rm th}$)	plotted as a function of neutron number. Gray-cross and blue-plus symbols respectively represent the residuals obtained from the bare and CatBoost-refined HFB masses with the corresponding Skyrme parameters.}
															      \label{Fig.17}
\end{figure}

{To further examine the extrapolation power of the CatBoost-refined HFB models with different Skyrme parameters for drip-line nuclei, similar to Ref.~\cite{Liu2025,Wu2020, Wu2024}, for each isotopic chain with $Z \geq 8$, the four most proton-rich and neutron-rich nuclei are respectively removed out from the training set, and according to the different extrapolation distances from the remain training set in the neutron direction, they are classified into four proton-rich and neutron-rich testing sets as illustrated in Fig.~\ref{Fig.16}(a). Note that for both proton-rich and neutron-rich sides, from the test set 1 to 4, the extrapolation distances are getting longer and longer. Except for the eight testing sets, the remaining nuclei are used for training the machine-learning model and the rms derivations of the binding-energy residuals between experiments and predictions are presented as functions of the extrapolation distance in Fig.~\ref{Fig.16}(b). One can see that the rms deviations of different Skyrme parameter sets are all very small on the training sets and smoothly increase with the increasing extrapolation distance, agreeing with the general trend as given, e.g., in Refs.~\cite{Liu2025,Wu2020, Wu2024}. Relatively, the prediction inaccurancies on the proton-rich side of the nuclear chart will grow more rapidly. It can still be noticed that even at large extrapolation distances except (e.g., see $\pi_4$), the CatBoost-refined HFB predictions are often better than those of the bare HFB models, possessing the extrapolation power for nuclei.       

In the superheavy region, relative to nuclear masses, there are more measured data of the $\alpha$-decay energies $Q_\alpha$. It may be instructive to evaluate the predictive power of the models based on $Q_\alpha$ energies owing to the the relationship between the $Q_\alpha (Z,N)$ value and nuclear mass excess $M (Z, N)$ or binding energy $B (Z, N)$ [namely, $Q_\alpha (Z,N)$ = $M (Z, N) - M (Z-2, N-2) -M (2, 2) $ =   $B (Z-2, N-2) + B(2, 2) -B (Z, N)$]~\cite{Heenen2015}. Fig.~\ref{Fig.17} illustrates the calculated residuals  $\delta Q_\alpha$  between experimental data and theoretical calculations by the bare and Catboot-refined HFB models for nuclei with $N > 150$. One can see that the CatBoost-refined HFB calculations reproduce the $Q_\alpha$ energies slightly better, especially before $N = 170$. Combining with Fig.~11, it can be concluded that the bare HFB calculations can to some extent descripe the $Q_\alpha$ energies better but fail reproducing the binding energies, while the CatBoost-refined HFB models describle both of them better.
}  

\begin{table}
\caption{Six sets of coefficients (a$_v$, a$_s$, a$_c$, and c$_{\rm sym}$) of a modified Bethe-Weizs\"acker mass formula obtained by fitting the binding energies $B_{\rm ml}$ from the CatBoost-refined HFB calculations with different Skyrme parameter sets. All the coefficients are in units of MeV. See text for more details.}
\begin{ruledtabular}
\renewcommand
\arraystretch{1.5}
\centering
\begin{tabular}{lcccc}
Parameter &\   $a_v$  &\ $a_s$ \  &\ $a_c$ \  &\ $c_{\rm sym}$  \\
\hline
SkM*&-14.9742&16.0155&0.6890\ \ &26.1315 \\
SkP&-15.4875&17.9647&0.7107\ \ &29.2322 \\
SLy4&-15.6781&18.6314&0.7202\ \ &30.9569 \\
SV-min&-15.2836&17.1896&0.7008\ \ &27.9955 \\
UNEDF0&-15.0279&16.2981&0.6849\ \ &26.8918 \\
UNEDF1&-15.0758&16.4157&0.6890\ \ &26.7600 \\
\end{tabular}                                                             
	                                                             \label{table6}
\end{ruledtabular}
\end{table}
{

It is noteworthy that for accurately predicting the masses of drip-line nuclei and superheavy nuclei, the surface energy coefficients and symmetry-energy coefficients for the Skyrme parametrizations will play a crucial role~\cite{Jodon2016,Dutra2012}. Of course, the accurate masses, in turn, can be used to fine tune the Skyrme energy density functionals. To evaluate these coefficients based on the predicted binding energies $B_{ml}$ for all bound nuclei, we perform the least-squares fitting with a modified Bethe-Weizs\"acker mass formula, e.g., cf. Eq.(2) in Ref.~~\cite{Wang2010}. As shown in Table~\ref{table6}, the coefficients $a_v$, $a_s$, $a_c$, and $c_{sym}$, related respectively to the volume, surface, Coulomb and symmetry energies, are listed and may be helpful for further analyzing other nuclear properties and for proposing new Skyrme parametrizations in the future.
}

\section{Summary}
\label{sec:summary}

In this paper we evaluate the performance and predictive power of the Hartree–Fock–Bogoliubov mass models with six different Skyrme forces and the combinations with a CatBoost algorithm. Comparing with the available experimental data, we can obtained that different nucleon-nucleon interactions give rather different mass residuals (equivalently, the root-mean-square deviations), indicating the missing physics to some extent. The pattern of the residual distribution illustrates the possible existance of the heteroscedasticity and model bias. With the CatBoost-refined HFB mass models, both the predictive powers and, simultaneously, the heteroscedasticity and model bias are significantly improved for each Skyrme parameter set. In addition, it is found that the CatBoost-refined HFB mass models can possess good prediction accurancies and stabilities and may rearrange the predictive powers of the original HFB ones with different Skyrme parameters. For different nucleon-nucleon Skyrme effective interactions, the CatBoost algorithm can pick up the missing physics to some extent but exibit different model-repair abilities. {Moreover, in the CatBoost machine-leaning algorithm, the generalization ability (namely, model performance for the unseen data) on the extrapolation predictions is still worthy of reference (e.g., less than the bare theoretical predictions) though it decreases rapidly with the increasing distance from the training set and usually worse than that of the interpolation ones. Obviously, in the very extreme nuclear regions, e.g., far away from the training data, the extrapolation performance of the machine learning will be unreliable and need to be further optimized. Of course, it may be of interest to generalize the CatBoost algorithm to predict other nuclear properties (such as fission barriers, excited states) by considering some suitable features in the future. }

\section*{Conflict of Interest}

The authors declare that they have no known competing financial
interests or personal relationships that could have appeared to
influence the work reported in this paper.

\section{Acknowledgement}
This work was supported by the Natural Science Foundation of Henan Province (No. 252300421478) and the National Natural Science Foundation
of China (No.11975209, No. U2032211, and No. 12075287). Some of the
calculations were conducted at the National Supercomputing Center in
Zhengzhou.
\section*{References}
%
\end{document}